\documentclass[aps,prd,amsmath,floats,floatfix,twocolumn,
  superscriptaddress,nofootinbib]{revtex4-2}

\pdfoutput=1 
\usepackage{diagbox}
\usepackage{multirow}
\usepackage{braket}
\usepackage{amsfonts}
\usepackage{amsmath}
\usepackage{amssymb}
\usepackage{mathtools}
\usepackage{amsthm}
\usepackage{bm}
\usepackage{dcolumn}
\usepackage{epsfig}
\usepackage{commath}
\usepackage{graphicx}
\usepackage{graphics}
\usepackage[utf8]{inputenc} 
\usepackage{latexsym}
\usepackage{rotating}
\usepackage[dvipsnames]{xcolor}
\usepackage{mathrsfs}
\usepackage{microtype}
\usepackage{verbatim}
\usepackage{url}

\usepackage[caption=false]{subfig}
\usepackage[normalem]{ulem}
\usepackage{tikz}
\usetikzlibrary{decorations.pathmorphing,arrows.meta,calc}

\usepackage[breaklinks=true]{hyperref}
\hypersetup{
    colorlinks=true,
    linkcolor=NavyBlue,
    filecolor=Magenta,
    urlcolor=NavyBlue,
    citecolor=NavyBlue
}

\usepackage{yfonts}
\usepackage{float}
\usepackage{xspace} 
\usepackage[toc,page]{appendix}
\usepackage{siunitx}
\usepackage{array}
\usepackage{enumitem}
\usetikzlibrary{arrows.meta,positioning}

\newcommand{\be}{\begin{equation}}
\newcommand{\ee}{\end{equation}}

\newcommand{\RW}{{\mbox{\tiny RW}}}

\newcommand{\GR}{{\mbox{\tiny GR}}}

\newcommand{\ba}{\begin{align}}
\newcommand{\ea}{\end{align}}

\makeatletter
\newcommand*{\rom}[1]{\expandafter\@slowromancap\romannumeral #1@}
\makeatother
\allowdisplaybreaks

\AtBeginDocument{%
    \newwrite\bibnotes
    \def\bibnotesext{Notes.bib}
    \immediate\openout\bibnotes=\jobname\bibnotesext
    \immediate\write\bibnotes{@CONTROL{REVTEX41Control}}
    \immediate\write\bibnotes{@CONTROL{%
    apsrev41Control,author="08",editor="1",pages="1",title="0",year="1"}}
    \if@filesw
    \immediate\write\@auxout{\string\citation{apsrev41Control}}%
    \fi
}

\newcommand{\UIUC}{Illinois  Center  for  Advanced  Studies  of  the  Universe \&
Department of Physics, University of Illinois at Urbana-Champaign, Urbana, Illinois 61801, USA}
\newcommand{\CAL}{Theoretical Astrophysics 350-17, California Institute of Technology, Pasadena, CA 91125, USA}
\newcommand{\AEI}{Max Planck Institute for Gravitational Physics (Albert Einstein Institute), D-14476 Potsdam, Germany}

\begin{document}

    \title{Curvature Reconstruction and Tensor Polarizations in Isospectrality-Breaking Black-Hole Perturbations}

    \author{Pratik Wagle}
    \email{pratik.wagle@aei.mpg.de}
    \affiliation{\AEI}
%
    \author{Dongjun Li}
    \email{dongjun@illinois.edu}
    \affiliation{\UIUC}
    \author{Yanbei Chen}
    \affiliation{\CAL}
    \author{Nicol\'as Yunes}
    \affiliation{\UIUC}

    \date{\today}
    \begin{abstract}

Testing general relativity with the ringdown phase of gravitational waves requires connecting modifications of black-hole perturbation dynamics to observable signatures in detector strain. In this work, we study the relation between curvature perturbations, metric master variables, and detector tensor polarizations for nonrotating black holes in dynamical Chern-Simons gravity. In this theory the axial gravitational perturbation couples to a dynamical scalar field, while the polar sector remains GR-like at the order considered. We first construct a map for the perturbed Weyl scalar \(\delta\Psi_0\), expressing its even-parity part in terms of the ZM master function and its odd-parity part in terms of the RW master function together with the scalar radial perturbation. We also derive the leading asymptotic tensor polarizations \(h_+\) and \(h_\times\) directly from the metric perturbations. The resulting strain is reconstructed from the radiative gravitational master functions: the ZM function in the even sector and the Cunningham-Price-Moncrief-normalized odd function in the axial sector for dynamical Chern Simons gravity. No independent local algebraic term proportional to the scalar radial function appears in the leading tensor strain. These results separate curvature-level scalar dependence from tensor-channel scalar-frequency responses and provide the formal reconstruction basis for companion work on parity-split beating and effective ringdown frequencies.
    \end{abstract}
    \maketitle


\section{Introduction}
\label{sec:introduction}

The ringdown of a perturbed black hole (BH) is one of the cleanest regimes in which gravitational-wave observations can be connected to linearized strong-field gravity. After merger, the remnant relaxes through quasinormal modes (QNMs), whose complex frequencies and damping times are determined by the perturbation equations of the final compact object. In general relativity (GR), the QNM spectrum of a Kerr BH is fixed by its mass and spin, making black-hole spectroscopy a direct test of the Kerr hypothesis and of the dynamics of GR in the strong-field regime~\cite{Dreyer:2003bv,Berti:2009kk,Berti:2015itd,Cardoso:2016rao,Abbott:2021mjq,Cardoso:2019mqo,Cardoso:2018ptl,Berti:2005ys,LIGOScientific:2025wao,LIGOScientific:2026wpt}.

For nonrotating BHs in GR, the perturbation problem has an additional special structure. Metric perturbations separate into even-parity and odd-parity sectors, governed respectively by the Zerilli-Moncrief (ZM) and Regge-Wheeler (RW) equations~\cite{Regge:1957td,Zerilli:1971wd,Moncrief:1974am,Vishveshwara:1970cc,Vishveshwara:1970zz}. Although these two equations have different effective potentials, they possess exactly the same QNM frequencies~\cite{Chandrasekhar:1975zza}. This is the Schwarzschild isospectrality relation, whose standard interpretation is tied to the Darboux and Chandrasekhar transformations between metric-based and curvature-based descriptions of the perturbations~\cite{Chandrasekhar:1975nkd,Chandrasekhar_1983}.

Many theories beyond GR break this structure. Extra fields, higher-curvature interactions, or parity-violating couplings can modify the perturbation equations, couple sectors that are independent in GR, and lift the even-odd degeneracy~\cite{Pani:2009wy,Cano:2020cao,Cano:2023jbk,Li:2022pcy,Hussain:2022ins,Chung:2024ira}. Recent work using the metric perturbation approach and the modified Teukolsky formalisms has provided a systematic route for computing ringdown corrections in such settings, including non-Ricci-flat or non-Petrov-type-D backgrounds, rotating dynamical Chern-Simons (dCS) BHs, and explicit studies of isospectrality breaking~\cite{Wagle:2021tam,Srivastava:2021imr,Li:2022pcy,Hussain:2022ins,Cano:2023tmv,Wagle:2023fwl,Li:2025fci,Aly:2026mtf,Molina:2010fb,Yunes:2007ss,Li:2023ulk,Alapati:2025jmd,Chung:2023wkd,Chung:2023zdq,Chung:2024ira,Chung:2025gyg,Cano:2024jkd}. These developments make it increasingly important to distinguish changes in the spectrum from changes in the observable waveform.

This distinction is especially relevant in dCS gravity. In the nonrotating limit, the background metric is Schwarzschild and the background scalar vanishes~\cite{Jackiw:2003pm,Grumiller:2007rv}, but the axial gravitational perturbation couples to the scalar perturbation~\cite{Molina:2010fb,Wagle:2021tam}. The polar gravitational perturbation remains GR-like at the order considered here~\cite{Yunes:2007ss,Cardoso:2009pk}. Consequently, the odd gravitational-led spectrum is shifted relative to the even gravitational-led spectrum, and the coupled odd system also contains scalar-led branches~\cite{Molina:2010fb,Pani:2011xj,Wagle:2021tam,Li:2025fci}. The existence of these propagating scalar degree of freedom, however, does not by itself determine what appears in the detector tensor strain.

Recent analyses of beyond-GR ringdown modeling have emphasized that extra propagating fields can produce more than shifted gravitational QNM frequencies~\cite{Chung:2024ira}. In particular, modified Teukolsky equations can lead to additional metric-response terms carrying the characteristic frequencies of the extra fields~\cite{Lestingi:2025jyb}. Related waveform-level studies have advocated ringdown templates containing both shifted gravitational-led modes and extra-field-induced modes, including applications to dCS and scalar-Gauss-Bonnet gravity~\cite{Crescimbeni:2024qca,Crescimbeni:2025kxi,Carullo:2021dui}. These works motivate a sharper reconstruction question: when a scalar appears in the perturbation equations or in a curvature variable, does it also appear as a direct local contribution to \(h_+\) and \(h_\times\), or only indirectly through the metric response?

The sensitivity of the LIGO-Virgo-KAGRA network has turned black-hole spectroscopy from a proof of principle into a precision measurement: the ringdown of GW250114 already constrains the dominant quadrupolar frequency to within a few percent of the Kerr prediction and tests the spectroscopic pattern involving the first overtone at the tens-of-percent level~\cite{LIGOScientific:2025wao}, and the fourth Gravitational-Wave Transient Catalog extends tests of the remnant across dozens of events~\cite{LIGOScientific:2026wpt}. In parallel, theory-agnostic searches for non-gravitational ringdown modes~\cite{Crescimbeni:2024qca,Crescimbeni:2025kxi} and parametrized frameworks for extra-field frequency contamination~\cite{Lestingi:2025jyb,Carullo:2021dui} are being developed specifically to interpret this kind of data. Every such search assumes, implicitly, some mapping between the bulk content of a modified-gravity perturbation problem (i.e.~the propagating scalar, vector and tensor degrees of freedom that are excited during black hole merger) and the tensor polarizations that can actually be detected, such as  \(h_+\) and \(h_\times\). Making that mapping explicit, rather than assumed, is a prerequisite for correctly attributing, or correctly excluding, a beyond-GR signature in the ringdown of a real event, and for avoiding the conflation of curvature-level bulk content with a directly measurable strain contamination.

The goal of this paper is to answer this reconstruction question for nonrotating dCS gravity. We choose to work in the nonrotating limit because it allows us to track the problem analytically. Historically, dCS perturbation theory developed in this order: the coupled axial-scalar equations and their shifted quasinormal spectrum were first derived for a nonrotating, Schwarzschild background~\cite{Yunes:2007ss,Cardoso:2009pk,Molina:2010fb}, and only later extended to slow rotation, either by direct perturbative solutions~\cite{Wagle:2021tam,Srivastava:2021imr} or through modified Teukolsky and spectral frameworks~\cite{Li:2022pcy,Wagle:2023fwl,Li:2025fci,Chung:2024ira,Chung:2025gyg}. 
The problem of reconstructing the metric perturbation and detector observables from master variables follows the same order for a more fundamental reason: the RW and ZM master variables do not exist once rotation is turned on. 

The above reconstruction problem is well known in GR. A Kerr background is only axisymmetric, so a metric perturbation mixes multipoles and admits no analogue of the Schwarzschild parity decomposition. The Teukolsky equation, instead, separates because of the Petrov type-D structure of the Kerr metric~\cite{Teukolsky:1973ha}, and recovering a metric perturbation from its curvature solution requires the gauge-dependent Chrzanowski-Cohen-Kegeles procedure~\cite{Chrzanowski:1975wv,Cohen_Kegeles_1975,Kegeles:1979an,Wald:1978vm,Ori:2002uv,Keidl:2006wk,Keidl:2010pm}. No such procedure has been developed yet in modified gravity, although work in this direction is rapidly progressing~\cite{Li:2026rkf}. 
Our focus on a nonrotating background is thus not merely a simple warm-up, but rather, it is the only setting in which a dictionary of this type can currently be written down.

We pursue the goal of this paper by separating three logically distinct levels of the calculation: the coupled bulk axial-scalar master equations, the curvature map from the master variables to \(\delta\Psi_0\) and back again, and the leading tensor strain that can be detected at future null infinity. We construct an explicit forward reconstruction hierarchy connecting these levels, of the schematic form
\begin{equation}
\begin{aligned}
\text{bulk axial-scalar system}
\;\Bigg\{
\begin{array}{l}
\longrightarrow\;\delta\Psi_0(\Psi_{\rm ZM},\Psi_{\rm RW},R_{\ell m})\\[0.4em]
\longrightarrow\;h^\infty(\Psi_{\rm ZM},\Psi^{\rm rad}_{O})
\end{array}
\end{aligned}
\label{eq:intro_reconstruction_hierarchy}
\end{equation}
rather than simply asserting that such a relation exists.
The main result of this paper is a Schwarzschild RW/ZM metric-master dictionary that is parity resolved, gauge independent, and invertible for \(\ell\ge2\) allowing us to directly connect perturbation variables to detector observables. 

In deriving and developing these three logically-distinct levels of reconstruction, we obtain several intermediate results. We show that the scalar perturbation enters the odd-parity Chandrasekhar map for \(\delta\Psi_0\) and \(\delta\Psi_4\) Weyl scalar perturbations explicitly. We then show that this curvature-level scalar dependence does not become an independent local scalar term in the leading tensor polarizations. Instead, the asymptotic metric perturbation strain is reconstructed entirely from the gravitational radiative master functions: the even contribution is controlled by the ZM function, while the odd contribution is controlled by the Cunningham-Price-Moncrief (CPM)-normalized radiative odd variable~\cite{Cunningham:1978zfa,Cunningham:1979px}.

This separation leads to a conclusion that is easy to state but not obvious in advance: explicit scalar dependence in a curvature-level quantity is \textit{not}, by itself, evidence for a direct scalar term in the measured polarizations. Polarization studies of dCS gravity have shown that gravitational waves admit only the + and x tensor polarization modes~\cite{Wagle:2019mdq}. Moreover, the leading metric perturbation tensor strain contains no independent local algebraic contribution from the scalar radial function. Any scalar-frequency content in \(h_+\) and \(h_\times\) can thus only enter \textit{indirectly}, through the effects of the scalar induced on the odd-parity metric response and its radiative residue. 
The phenomenological consequences of these results, including the interaction between parity-split modes, the conditions under which a scalar-led frequency becomes observable, and the associated effective-frequency of the ringdown strain, are developed in a companion paper~\cite{paper2}.

Determining the correct radiative residue is not automatic. The RW variable natural to the axial field equation and the CPM-normalized variable entering the radiative strain share the same QNM pole locations but different outgoing residues, related by a frequency-dependent normalization~\cite{Nagar:2005ea}. An analogous normalization subtlety arises between field-equation and radiative variables in the rotating, Teukolsky-based setting~\cite{Sasaki:1981sx}. This distinction between field-equation and radiative normalization is built into our explicit reconstruction of \(h_+\) and \(h_\times\). Our observable-strain reconstruction is carried out in the CPM normalization, and it is precisely this choice that yields the clean result above: the leading odd-parity strain is controlled solely by \(\Psi^{\rm rad}_O\equiv\Psi_{\rm CPM}\), with no separate scalar-perturbation dependent term. This distinction is immaterial if one only asks for the spectrum, but it is exactly what our reconstruction depends on to correctly state whether a scalar-led frequency survives in the detector strain.

The paper is organized as follows. Sec.~\ref{sec:review} reviews the metric perturbation setup, the GR RW and ZM equations, and the dCS axial-scalar-polar perturbation system. Sec.~\ref{sec:chandra_dcs_schw} constructs the nonrotating Chandrasekhar map for \(\delta\Psi_0\) and verifies its consistency with the modified Teukolsky equation. Sec.~\ref{sec:grobs} presents the GR asymptotic polarizations with explicit attention to the odd-sector radiative normalization. Sec.~\ref{sec:dcsobs} extends the metric polarizations derivation to dCS gravity and proves that the leading tensor polarizations contain no independent local scalar term. Sec.~\ref{sec:conclusion} summarizes the formal implications and identifies the residue-level questions deferred to companion work~\cite{paper2}. 

Henceforth, we use the following conventions unless stated otherwise. We work in 4-dimensions with metric signature $(-,+,+,+)$ as in~\cite{Misner:1973prb}. The spacetime coordinates are denoted with Greek symbols $(\mu, \nu, \ldots)$, while temporo-radial coordinates are represented with upper-case Latin letters $(A, B, \ldots)$ and angular coordinates with lower-case Latin letters $(a,b,\ldots)$. 
We also use geometric units and set $G=c=1$. We set the Fourier convention \(e^{-i\omega t}\), and the four-dimensional orientation fixed by \(\epsilon^{tr\theta\phi}=+1/\sqrt{-g}\). These choices, together with \(\epsilon_{\theta\phi}=-\sin\theta\) on the unit two-sphere, fix the relative signs of the dCS scalar source and the odd-parity strain phase. A different orientation or axial-harmonic convention changes the corresponding overall odd-sector sign but not the physical pole content. Pole locations and residues throughout refer to the standard QNM boundary conditions, purely ingoing at the future horizon and purely outgoing at spatial infinity. 

\section{BH perturbation theory in GR and beyond} \label{sec:review}

In this section, we review the perturbative ingredients needed for the reconstruction calculation: the RW~\cite{Regge:1957td} and ZM~\cite{Zerilli:1971wd,Moncrief:1974am} equations, the Teukolsky formalism~\cite{Teukolsky:1973ha}, and its modified extensions~\cite{Wagle:2021tam,Srivastava:2021imr,Li:2022pcy,Hussain:2022ins,Cano:2023tmv}. Readers already familiar with these topics may wish to skip ahead to Sec.~\ref{sec:chandra_dcs_schw}.

\subsection{Metric perturbation approach} \label{sec:metpert}

We consider linear perturbations of a black-hole spacetime described by a stationary and spherically-symmetric background metric
$\bar{g}_{\mu\nu}$ and a small perturbation $h_{\mu\nu}$,
\begin{equation} \label{eq:perts}
    g_{\mu\nu} = \bar{g}_{\mu\nu} + \epsilon \, h_{\mu\nu}\,,
\end{equation}
where $\epsilon$ is a bookkeeping parameter. At linear order, $h_{\mu\nu}$ denotes a generic metric perturbation before any gauge choice is imposed.

For a spherically symmetric background, the spacetime can be decomposed as
$\mathcal{M} = \mathcal{M}_2 \times S^2$, where $\mathcal{M}_2$ is spanned by $x^A=(t,r)$ coordinates and
$S^2$ by $x^a=(\theta,\phi)$ coordinates. The perturbation $h_{\mu\nu}$ can then be expanded in tensor
spherical harmonics and decomposed into odd-parity (axial) and even-parity (polar) sectors,
\begin{equation}
h_{\mu\nu}
=
\sum_{\ell,m}
\left[
\left(h^{\ell m}_{\mu\nu}\right)^{(o)}
+
\left(h^{\ell m}_{\mu\nu}\right)^{(e)}
\right],
\end{equation}
where
\begin{equation}
\sum_{\ell,m}
\equiv
\sum_{\ell=2}^{\infty}
\sum_{m=-\ell}^{\ell}.
\end{equation}
The restriction to $\ell \geq 2$ reflects the fact that only these multipoles contribute to radiative gravitational degrees of freedom.
The odd-parity perturbation is written as
\begin{align}
\left(h_{\mu\nu}^{\ell m}\right)^{(o)}
=
\begin{pmatrix}
0 & h_{A}^{(o)}\, S_{c}^{\ell m} \\
h_{A}^{(o)}\, S_{c}^{\ell m} & 2h\, \nabla_{(d} S_{c)}^{\ell m}
\end{pmatrix},
\label{eq:hodd}
\end{align}
while the even-parity perturbation is written as
\begin{widetext}
\begin{equation} \label{eq:heven}
\left(h^{\ell m}_{\mu\nu}\right)^{(e)}
=
\left(
\begin{array}{cc|c}
H_{0}\, Y^{\ell m} & H_{1}\, Y^{\ell m} & h^{(e)}_{A}\, \nabla_{c} Y^{\ell m} \\[4pt]
H_{1}\, Y^{\ell m} & H_{2}\, Y^{\ell m} & {} \\ \hline
h^{(e)}_{A}\, \nabla_{c} Y^{\ell m} & {} &
r^{2}\!\left( K\,Y^{\ell m}\,\gamma_{cd} + G\,\nabla_{d}\nabla_{c} Y^{\ell m} \right)
\end{array}
\right),
\end{equation}
\end{widetext}
where $h_A^{(o)}$ and $h$ in the axial sector, and $H_{0}$, $H_{1}$, $H_{2}$, $h^{(e)}_{A}$, $K$, and $G$ in the polar sector are all functions of $(t,r)$ only. We suppress the $(\ell,m)$ labels on these functions for notational simplicity. The angular dependence of the axial metric perturbation is encoded in the axial vector spherical harmonics,
\begin{equation} \label{eq:vector_sph_harm}
    S^{\ell m}_{c} = \epsilon_{cd}\gamma^{de}\nabla_e Y^{\ell m}(\theta, \phi) \,,
\end{equation}
where $\nabla_c$ is the covariant derivative compatible with the metric on the unit two-sphere, $\gamma_{cd} = \mathrm{diag}(1,\sin^2\theta)$, and $\epsilon_{cd}$ is the Levi-Civita tensor on $S^2$, satisfying $\nabla_a \epsilon_{bc}=0$.
We use unit-weight symmetrization, \(A_{(ab)}=(A_{ab}+A_{ba})/2\), and choose the two-sphere orientation \(\epsilon_{\theta\phi}=-\sin\theta\). With this convention \(S_\theta=-(\sin\theta)^{-1}\partial_\phi Y^{\ell m}\) and \(S_\phi=\sin\theta\,\partial_\theta Y^{\ell m}\). Reversing the axial-harmonic orientation changes the overall odd-parity phase convention, but not the pole content of the waveform. Here and throughout, $Y^{\ell m}$ denotes the scalar spherical harmonics.

For the linear perturbations considered in this work, only the usual tensorial \( + \) and \(\times\) polarizations of the metric perturbation survive at spatial infinity, and thus, reach gravitational-wave detectors, as shown explicitly for the dCS and sGB settings relevant here~\cite{Sopuerta:2009iy,Wagle:2019mdq}. We therefore restrict the asymptotic strain to these tensor modes. For a spherically symmetric background, one can further exploit the gauge freedom and work in the RW gauge~\cite{Regge:1957td}. In this gauge, the perturbation amplitudes $h^{(e)}_{0}$, $h^{(e)}_{1}$, $h$ and $G$ appearing in Eqs.~\eqref{eq:hodd} and \eqref{eq:heven} can be set to zero. This gauge choice is particularly useful because it removes redundant variables and allows the perturbation equations to be recast in terms of a reduced set of gauge-fixed amplitudes or, equivalently, a smaller number of master variables.

\subsection{Master equations in GR} \label{sec:RWZM-GR}

In this subsection, we briefly review the derivation of master equations in GR, wherein for a nonrotating background, the two parity sectors decouple exactly. Rotation generically introduces couplings between different multipoles and parity sectors at the level of the perturbed field equations~\cite{Pani:2013pma,Teukolsky:1973ha,Leaver:1985ax,Teukolsky:1974yv}, but this will not be considered here.

In GR, perturbations of a Schwarzschild BH are governed by the vacuum Einstein equations,
\begin{equation}
G_{\mu\nu}[g]=0,
\end{equation}
with the metric expanded as in Eq.~\eqref{eq:perts}. Linearizing about the Schwarzschild background,
\begin{equation} \label{eq:schw}
ds^2=-f(r)\,dt^2+f(r)^{-1}dr^2+r^2 d\Omega^2,
\qquad
f(r)=1-\frac{2M}{r},
\end{equation}
one obtains the first-order field equations
\begin{equation}
\delta G_{\mu\nu}[h]=0.
\end{equation}
After decomposing the perturbation into tensor spherical harmonics, the odd- and even-parity sectors decouple for the nonrotating background and can be treated independently, as shown below.
For convenience we define,
\begin{equation}
\Lambda \equiv \ell(\ell+1),
\qquad
\lambda \equiv \Lambda-2 .
\label{eq:lambda_def}
\end{equation}

\subsubsection{Odd parity master function and the Regge-Wheeler equation} \label{sec:RWGR}

In the odd-parity sector, the metric perturbation is described by the amplitudes $h_A^{(o)}=(h_0,h_1)$ and $h$ introduced in Eq.~\eqref{eq:hodd}. It is often convenient to define
\begin{equation}
h_2 \equiv 2h.
\end{equation}
This convention is used whenever the generic-gauge angular amplitude is denoted by \(h_2\)~\cite{Nagar:2005ea}. The observable-strain formulas below are written in terms of the metric amplitude \(h\) of Eq.~\eqref{eq:hodd} and therefore any equation solved for \(h_2\) must be divided by two before it is converted into \(h\).
Substituting the odd-parity perturbation ansatz into the linearized vacuum Einstein equations and projecting onto the axial tensor spherical harmonics yields a coupled system of equations for $h_0$, $h_1$, and $h_2$. This system can be reduced to a single decoupled second-order equation for a suitably chosen master variable.

Without imposing any gauge condition, the odd-parity gauge-invariant master function can be written as~\cite{Regge:1957td,Pani:2013pma,Maggiore:2018sht}
\begin{equation} \label{eq:RW_general_gauge}
\Psi^{(o)}_{\ell m}(t,r)
=
\frac{f(r)}{r}
\left[
h_{1}^{\ell m}(t,r)
-
\frac{r^{2}}{2}\,
\partial_r\!\left(\frac{h_{2}^{\ell m}(t,r)}{r^{2}}\right)
\right].
\end{equation}
This combination is invariant under infinitesimal odd-parity gauge transformations at linear order and therefore captures the physical axial gravitational perturbation independently of the coordinate choice. The advantage of this form is that the RW equation can be derived directly from the generic odd-parity perturbation, without first fixing the RW gauge.

Introducing the tortoise coordinate $r_*$ through
\begin{equation}
\frac{dr_*}{dr}=f(r)^{-1},
\label{eq:tortoise-coord}
\end{equation}
the gauge-invariant master variable satisfies the RW equation,
\begin{equation} \label{eq:RW_time_domain}
\left[
-\frac{\partial^2}{\partial t^2}
+
\frac{\partial^2}{\partial r_*^2}
-
V_{\rm RW}(r)
\right]
\Psi^{(o)}_{\ell m}(t,r)=0,
\end{equation}
where the RW potential is
\begin{equation} \label{eq:VRW}
V_{\rm RW}(r)
=
f(r)
\left[
\frac{\ell(\ell+1)}{r^2}
-
\frac{6M}{r^3}
\right].
\end{equation}
Upon a Fourier decomposition,
\begin{equation}
\Psi^{(o)}_{\ell m}(t,r)=\Psi^{(o)}_{\ell m}(r)e^{-i\omega t},
\end{equation}
Eq.~\eqref{eq:RW_time_domain} becomes
\begin{equation} \label{eq:RW_freq_domain}
\frac{d^2 \Psi^{(o)}_{\ell m}}{dr_*^2}
+
\left[
\omega^2-V_{\rm RW}(r)
\right]
\Psi^{(o)}_{\ell m}=0.
\end{equation}

If one instead imposes the RW gauge from the outset, then the odd-parity amplitude $h_2^{\ell m}$ is set to zero. In that case, the master function in Eq.~\eqref{eq:RW_general_gauge} simplifies to
\begin{equation} \label{eq:RW_RWgauge}
\Psi^{(o)}_{\ell m}(t,r)
=
\frac{f(r)}{r}\,h_1^{\ell m}(t,r).
\end{equation}
The remaining odd-parity field equations can then be combined directly into the same decoupled wave equation, Eqs.~\eqref{eq:RW_time_domain} and \eqref{eq:RW_freq_domain}. Thus, the use of the RW gauge simplifies the algebra, but does not alter the final master equation or the associated effective potential. 
The functions in Eqs.~\eqref{eq:RW_general_gauge} and~\eqref{eq:RW_RWgauge} are commonly written as the RW function \(\Psi_{\rm RW}\), without and within the RW gauge respectively.
In the field-equation and Chandrasekhar-map sections, Secs.~\ref{sec:RWGR}--\ref{sec:oddchandra}, we denote the RW variable by \(\Psi_{\rm RW}\). 
In the observable-strain sections, Secs.~\ref{sec:odd_obs_gr} and \ref{sec:dcsobs_odd}, the relevant radiative odd variable is instead the CPM-normalized variable, denoted by \(\Psi^{\rm rad}_{O}\equiv\Psi_{\rm CPM}\) [c.f. Eq.~\eqref{eq:rw_cpm_freq_dictionary}]. 
The CPM master function is defined as 
\begin{align}
\Psi_{{\rm CPM},\ell m}(t,r)
=
\frac{r}{\lambda}
\left[
\partial_t h_{1}^{\ell m}(t,r)
-\right. & \left.
\partial_r h_{0}^{\ell m}(t,r) \right.
\nonumber \\
& \left. \qquad + 
\frac{2}{r}h_{0}^{\ell m}(t,r)
\right].
\end{align}

The two variables have the same pole locations but different residues, related by a frequency-dependent normalization.

It is useful to distinguish between the gauge-invariant and gauge-fixed derivations of the RW equation. In the former, one begins with a generic odd-parity perturbation and identifies the combination in Eq.~\eqref{eq:RW_general_gauge}, which is invariant under infinitesimal gauge transformations and directly captures the physical axial degree of freedom. In the latter, one imposes the RW gauge, in which the same master variable reduces to the simpler form in Eq.~\eqref{eq:RW_RWgauge}. These two routes are entirely equivalent and lead to the same decoupled wave equation with the same effective potential. This equivalence is important because it makes clear that the RW equation governs the true dynamical content of the odd-parity sector rather than a gauge artifact. The distinction becomes useful when extending the formalism to more general settings in which a gauge-invariant treatment is preferable or gauge fixing does not simplify the system as strongly.

\subsubsection{Even-parity master function and Zerilli-Moncrief equation} \label{sec:ZMGR}

We now turn to the even-parity sector. As in the odd-parity case, substituting the even-parity perturbation ansatz of Eq.~\eqref{eq:heven} into the linearized Einstein equations and projecting onto the corresponding tensor spherical harmonics yields a coupled system of equations for the amplitudes $H_0$, $H_1$, $H_2$, $h^{(e)}_0$, $h^{(e)}_1$, $K$, and $G$. This system can again be reduced to a single decoupled master equation for a suitable gauge-invariant variable.

Following Moncrief~\cite{Moncrief:1974am}, one may introduce the gauge-invariant combinations
\begin{equation} \label{eq:k1}
k_{1}
\equiv
K
+
f(r)
\left(
r \, \partial_{r} G
-
\frac{2}{r} h_{1}^{(\mathrm{e})}
\right),
\end{equation}

\begin{equation} \label{eq:k2}
k_{2}
\equiv
\frac{1}{2}
\left[
f(r)\,H_{2}
-
\sqrt{f(r)}\,\partial_{r}\!\left(\frac{r\,K}{\sqrt{f(r)}}\right)
\right],
\end{equation}
where $f(r)$ is the Schwarzschild factor given in Eq.~\eqref{eq:schw}. Both $k_1$ and $k_2$ are invariant under infinitesimal even-parity gauge transformations at linear order. A further gauge-invariant combination is then defined as
\begin{equation} \label{eq:q1}
q_{1}
\equiv
r \Lambda k_{1}
+
4r\, f(r)\, k_{2}.
\end{equation}
where \(\Lambda\) is given by Eq.~\eqref{eq:lambda_def}.
The even-parity master function can then be written as
\begin{equation} \label{eq:Zerilli_master}
\Psi^{(\mathrm{e})}_{\ell m}(t,r)
=
\frac{r \, q_{1}}{\Lambda \bigl[ r(\Lambda - 2) + 6M \bigr]}.
\end{equation}
This master variable is gauge invariant and encodes the physical even-parity gravitational degree of freedom.

Introducing the tortoise coordinate through Eq.~\eqref{eq:tortoise-coord}, one finds that $\Psi^{(\mathrm{e})}_{\ell m}$ satisfies the ZM equation,
\begin{equation} \label{eq:Zerilli_time}
\left[
-\frac{\partial^2}{\partial t^2}
+
\frac{\partial^2}{\partial r_*^2}
-
V_{\ell}^{(\mathrm{e})}(r)
\right]
\Psi^{(\mathrm{e})}_{\ell m}(t,r)
=
0,
\end{equation}
in vacuum. The Zerilli potential is given by
\begin{align} \label{eq:Zerilli_potential}
V_{\ell}^{(\mathrm{e})}
= &
\frac{f(r)}{r^{3} \bigl[ (\Lambda - 2) r + 6M \bigr]^{2}}
 \bigl[\Lambda(\Lambda - 2)^{2} r^{3} \nonumber \\ 
& + 6(\Lambda - 2)^{2} M r^{2}  + 36(\Lambda - 2) M^{2} r + 72 M^{3}\bigr] \,.
\end{align}
Upon Fourier decomposition,
\begin{equation}
\Psi^{(\mathrm{e})}_{\ell m}(t,r)
=
\Psi^{(\mathrm{e})}_{\ell m}(r)e^{-i\omega t},
\end{equation}
the Zerilli equation reduces to
\begin{equation} \label{eq:Zerilli_freq}
\frac{d^2 \Psi^{(\mathrm{e})}_{\ell m}}{dr_*^2}
+
\left[
\omega^2
-
V_{\ell}^{(\mathrm{e})}(r)
\right]
\Psi^{(\mathrm{e})}_{\ell m}
=
0.
\end{equation}

In the RW gauge, one sets
\begin{equation}
h^{(e)}_0=h^{(e)}_1=G=0,
\end{equation}
so that the even-parity perturbation is described only by $H_0$, $H_1$, $H_2$, and $K$. In this gauge, the master function in Eq.~\eqref{eq:Zerilli_master} simplifies considerably and may be written directly in terms of the remaining metric amplitudes. A common form is
\begin{align} \label{eq:Zerilli_RWgauge}
\Psi^{(\mathrm{e})}_{\ell m}(t,r)
&=
\frac{r}{\Lambda}
\left[
K^{\ell m} \right. \nonumber \\
& \qquad  \left. +
\frac{2r}{\lambda r+6M}
\Bigl(
f(r)^{2}H_2^{\ell m}
-
r f(r)\,\partial_r K^{\ell m}
\Bigr)
\right].
\end{align}
or equivalently in algebraically rearranged forms used in the literature. 
In the RW gauge, the even parity master function, usually called the ZM function, is expressed as~\cite{Zerilli:1971wd,Moncrief:1974am,Berti:2009kk,Maggiore:2018sht} 
\begin{equation}
\Psi^{\ell m}_{\rm ZM}(r)
=
\frac{
r^{2}K^{\ell m}(r)
-\dfrac{i(r-2M)}{\omega}H^{\ell m}_{1}(r)
}{
6M+\lambda r
}\,.
\end{equation}
As in the odd-parity case, the gauge-fixed and gauge-invariant derivations are entirely equivalent and lead to the same master equation.

The Zerilli equation therefore plays for the even-parity sector the same role that the RW equation plays for the odd-parity sector: it isolates the physical dynamical degree of freedom and recasts the linearized Einstein equations into a one-dimensional wave equation in an effective potential. This formulation is the natural starting point for the study of polar quasinormal modes, waveform reconstruction, and the relation between even- and odd-parity perturbations in Schwarzschild spacetime.

\subsection{Dynamical Chern-Simons gravity: theory and perturbations} \label{sec:RWZM-bGR}

We now turn to modified theories of gravity, with particular emphasis on dCS gravity. In such theories, the Einstein-Hilbert action is supplemented by additional fields and non-minimal curvature couplings, which modify the perturbation equations and, in general, introduce new dynamical degrees of freedom. In dCS gravity, the action is~\cite{Jackiw:2003pm,Alexander:2009tp}
\begin{equation} \label{eq:Action1}
S = S_{\rm EH} + S_{\vartheta} + S_{\rm CS},
\end{equation}
where the Einstein-Hilbert term is
\begin{equation} \label{eq:EH}
S_{\rm EH} = \kappa \int d^4x \sqrt{-g}\, R,
\end{equation}
with $\kappa=(16\pi)^{-1}$. The scalar-field contribution is
\begin{equation} \label{eq:sf1}
S_{\vartheta}
=
-\frac{1}{2}\int d^4x \sqrt{-g}
\left[
g^{\mu\nu}(\nabla_\mu \vartheta)(\nabla_\nu \vartheta)
+2V(\vartheta)
\right],
\end{equation}
where $\nabla_\mu$ is the covariant derivative compatible with $g_{\mu\nu}$. In what follows, we set the scalar potential to zero, $V(\vartheta)=0$. We have also rescaled the scalar so that the usual kinetic coupling is \(\beta=1\). The dCS interaction term is
\begin{equation} \label{eq:SCS}
S_{\rm CS}
=
\frac{\alpha}{4}\int d^4x \sqrt{-g}\,\vartheta\, {}^\ast\!RR,
\end{equation}
where $\alpha$ is the dCS coupling constant, with dimensions of length squared, and
\begin{equation}
{}^\ast\!RR
=
{}^\ast\!R^{\mu}{}_{\nu}{}^{\rho\sigma}
R^{\nu}{}_{\mu\rho\sigma}
\end{equation}
is the Pontryagin density. Here,
\begin{equation}
{}^\ast\!R^{\mu}{}_{\nu}{}^{\rho\sigma}
=
\frac{1}{2}
\epsilon^{\rho\sigma\alpha\beta}
R^{\mu}{}_{\nu\alpha\beta},
\end{equation}
with $\epsilon^{\mu\nu\rho\sigma}$ the Levi-Civita tensor.

Varying the action with respect to the inverse metric and the scalar field yields the dCS field equations. Variation with respect to $g^{\mu\nu}$ gives
\begin{equation} \label{eq:field-eq}
G_{\mu\nu}
+
\frac{\alpha}{\kappa} C_{\mu\nu}
=
\frac{1}{2\kappa} T_{\mu\nu}^{\vartheta},
\end{equation}
where $G_{\mu\nu}$ is the Einstein tensor,
\begin{equation} 
\label{eq:CTensor}
C_{\mu\nu}
=
(\nabla_\sigma \vartheta)\,
\epsilon^{\sigma\delta\alpha}{}_{(\mu}
\nabla_\alpha R_{\nu)\delta}
+
(\nabla_\sigma \nabla_\delta \vartheta)\,
{}^\ast\!R^{\delta}{}_{(\mu\nu)}{}^{\sigma}
\end{equation}
is the dCS correction tensor, and
\begin{equation} \label{eq:SETfield}
T_{\mu\nu}^{\vartheta}
=
(\nabla_\mu \vartheta)(\nabla_\nu \vartheta)
-
\frac{1}{2}g_{\mu\nu}
(\nabla^\sigma \vartheta)(\nabla_\sigma \vartheta)
\end{equation}
is the scalar-field stress-energy tensor. Variation with respect to $\vartheta$ gives the sourced scalar wave equation,
\begin{equation} \label{eq:CSfield}
\Box \vartheta
=
-\frac{\alpha}{4}\, {}^\ast\!RR.
\end{equation}

We now perturb both the metric and the scalar field about a stationary and spherically symmetric BH background. One can show that the Schwarzschild metric continues to be the correct spacetime to describe non-rotating BHs in dCS gravity~\cite{Alexander:2009tp}. Using the metric decomposition in Eqs.~\eqref{eq:perts}, \eqref{eq:hodd}, and \eqref{eq:heven}, together with a decomposition of the scalar perturbation into spherical harmonics as,
\begin{equation} \label{eq:sfperts}
    \delta\vartheta(t,r,\theta,\phi)
    =
    \sum_{\ell\ge0,m}
    \frac{R_{\ell m}(t,r)}{r}
    Y_{\ell m}(\theta,\phi)\,,
\end{equation}
and then focusing on a fixed \((\ell,m)\) mode, one finds that the perturbation equations can again be organized into odd- and even-parity sectors. The explicit coefficients displayed below use the scalar normalization in Eq.~\eqref{eq:sfperts}. Scalar perturbations can in principle also have \(\ell=0,1\) components; the restriction \(\ell\ge2\) applies only to the detector tensor-radiative channels. In dCS gravity, however, the scalar field couples only to the odd-parity gravitational sector, while the even-parity gravitational sector remains decoupled at the order relevant here. As a result, the odd-parity gravitational perturbation and the scalar perturbation satisfy a coupled system of master equations, whereas the even-parity perturbation continues to satisfy a homogeneous equation.

Working with a nonrotating BH background metric in dCS gravity, and to quadratic order in the coupling \(\alpha\), the resulting frequency-domain master equations were derived in Refs.~\cite{Molina:2010fb}. They can be written as
\begin{widetext}
\begin{subequations}
\label{eqs:dcsperturbed_nonrot}
\begin{align}
\label{eq:SF}
f(r)^2 \partial_{rr}R_{\ell m}
+ \frac{2M}{r^2} f(r)\, \partial_r R_{\ell m}
+ \left[\omega^2 - V^{S}_{\rm eff}(r,\alpha^2)\right] R_{\ell m}
&= -\,f(r)\,
\frac{6i\,\alpha\,(\ell-1)\ell(\ell+1)(\ell+2)\,M}{r^5 \omega }\,
\Psi^{\rm RW}_{\ell m} \, , \\
\label{eq:RW}
f(r)^2 \partial_{rr}\Psi^{\rm RW}_{\ell m}
+ \frac{2M}{r^2} f(r)\, \partial_r \Psi^{\rm RW}_{\ell m}
+ \left[\omega^2 - V^{A}_{\rm eff}(r)\right]\Psi^{\rm RW}_{\ell m}
&= \,f(r)\,
\frac{6i\,\alpha\,M\,\omega}{\kappa\, r^5}\,
R_{\ell m} \, , \\
\label{eq:ZM}
f(r)^2 \partial_{rr}\Psi^{\rm ZM}_{\ell m}
+ \frac{2M}{r^2} f(r)\, \partial_r \Psi^{\rm ZM}_{\ell m}
+ \left[\omega^2 - V^{P}_{\rm eff}(r)\right]\Psi^{\rm ZM}_{\ell m}
&= 0 \, .
\end{align}
\end{subequations}
\end{widetext}
Here, $\Psi^{\rm RW}_{\ell m}$ denotes the dCS-modified odd-parity gravitational master function, $R_{\ell m}$ is the radial scalar perturbation, and $\Psi^{\rm ZM}_{\ell m}$ denotes the even-parity master function, which coincides with the GR ZM variable at the nonrotating order retained here. At nonrotating order, the RW and ZM master variables depend on the odd- and even-parity metric perturbation components, respectively, through exactly the same combinations as in GR, as reviewed in Sec.~\ref{sec:review}.
The functions $V_{\rm eff}^{S}$, $V_{\rm eff}^{A}$, and $V_{\rm eff}^{P}$ are the effective scalar, axial, and polar potentials, respectively, and are given by
\begin{subequations}
\label{eqs:potentials_nonrot}
\begin{align}
&V^{S}_{\rm eff}(r,\alpha^2)
= f(r)\left[\frac{\ell(\ell+1)}{r^2}
\left(1+\frac{576\pi M^2\alpha^2}{r^6}\right)
+\frac{2M}{r^3}\right] \, , \\
&V^{A}_{\rm eff}(r)
= f(r)\left[\frac{\ell(\ell+1)}{r^2}-\frac{6M}{r^3}\right] \, , \\
&V^{P}_{\rm eff}(r) = V^{(\mathrm{e})}_{\ell}(r) \,,
\end{align}
\end{subequations} 
where $V^{(\mathrm{e})}_{\ell}(r)$ is given by Eq.~\eqref{eq:Zerilli_potential}. The scalar equation, Eq.~\eqref{eq:SF}, contains an explicit factor of \(1/\omega\). The displayed master system therefore applies to nonzero-frequency modes and static and zero-frequency perturbations require separate treatment.

The structure of Eqs.~\eqref{eq:SF}--\eqref{eq:ZM} is physically important. The scalar perturbation couples only to the odd-parity gravitational sector, and this coupling persists already in the nonrotating limit. By contrast, the even-parity sector remains decoupled and obeys a homogeneous equation. In the strictly nonrotating dCS case, the polar gravitational sector is the GR Zerilli sector at the order considered here. Corrections to the polar potential arise only when additional effects, such as spin corrections or higher-order modifications, are retained. This separation of sectors underlies the parity-dependent ringdown phenomenology discussed later in this work.

\section{Chandrasekhar transformation for nonrotating dCS gravity}
\label{sec:chandra_dcs_schw}

In BH perturbation theory, one often encounters two equivalent but technically distinct descriptions of the linear dynamics. On the one hand, there are Teukolsky-type equations, written in terms of curvature perturbations, whose effective potentials are generally long-ranged~\cite{Teukolsky:1973ha,Teukolsky:1974yv}. On the other hand, there are RW/ZM type equations~\cite{Regge:1957td,Zerilli:1971wd,Moncrief:1974am}, written in terms of metric-based master variables, with short-ranged potentials and simpler asymptotic behavior. In GR, Chandrasekhar showed that, for a Schwarzschild BH, these two descriptions are related by an explicit differential transformation connecting the Teukolsky equation for Weyl scalars to the RW and ZM equations for odd- and even-parity metric perturbations, respectively~\cite{Chandrasekhar:1975nkd,Chandrasekhar_1983}. In this section, we construct the corresponding Chandrasekhar transformation in the nonrotating limit of dCS gravity and describe how it can be implemented in practice.

In the nonrotating limit, the background BH solution in dCS gravity coincides with the Schwarzschild metric, while the theory contains in addition a dynamical scalar degree of freedom. At linear order, the parity structure of the perturbations implies that the odd-parity gravitational sector couples to the scalar perturbation, whereas the even-parity sector remains decoupled, as seen from Eqs.~\eqref{eqs:dcsperturbed_nonrot}. Consequently, in dCS gravity the Chandrasekhar transformation relates two descriptions of the coupled gravitational-scalar perturbation system, rather than two independent second-order equations. Recasting the system in terms of short-ranged master variables with standard ingoing and outgoing asymptotic behavior is useful for QNM and scattering calculations.

More concretely, the strategy is to evaluate the Weyl curvature scalar on a perturbed background, decompose the metric perturbations into RW and ZM master functions and then use the RW and ZM master equations given in Eqs.~\eqref{eqs:dcsperturbed_nonrot} to eliminate higher radial derivatives. In the odd-parity sector, this procedure also generates explicit dependence on the scalar perturbation through the dCS coupling. The final result is a transformation expressing the curvature perturbation in terms of the metric master variables and the scalar mode. Although we focus here on dCS gravity, the same logic applies more broadly: once the appropriate master equations are known, the Chandrasekhar transformation can be constructed systematically for other modified gravity theories as well.

The construction proceeds as follows:
\begin{enumerate}
\item choose a background null tetrad $l^{\mu}$, $n^{\mu}$, $m^{\mu}$, $\bar{m}^{\mu}$ for the nonrotating BH in the modified theory and work within the RW gauge; 
\item compute $\Psi_0=C_{\mu \nu \rho \sigma}\,l^{\mu} m^{\nu} l^{\rho} m^{\sigma}$, where $C_{\mu \nu \rho \sigma}$ is the Weyl tensor, and linearize it in the perturbations to find the linearized $\delta \Psi_0$;
\item decompose the result into even- and odd-parity metric perturbation sectors and perform the harmonic decomposition;
\item rewrite the even-parity radial piece in terms of $\Psi_{\rm ZM}(r)$ and $\Psi_{\rm ZM}'(r)$, and eliminate $\Psi_{\rm ZM}''(r)$ using the polar master equation in Eq.~\eqref{eq:ZM};
\item rewrite the odd-parity radial piece in terms of $\Psi_{\rm RW}(r)$ and $\Psi_{\rm RW}'(r)$, and eliminate $\Psi_{\rm RW}''(r)$ using the axial master equation in Eq.~\eqref{eq:RW}, thereby generating a scalar contribution proportional to $R_{\ell m}(r)$;
\item assemble the full nonrotating transformation
\begin{equation}
\delta\Psi_0
=
\delta\Psi_0\bigl(\Psi_{\rm ZM},\Psi_{\rm ZM}',\Psi_{\rm RW},\Psi_{\rm RW}',R_{\ell m}\bigr);
\end{equation}
\item verify by direct substitution that the transformed $\delta \Psi_0$  satisfies the modified Teukolsky equation. At first order in the metric perturbation about a Schwarzschild background, \(\delta\Psi_0\) is invariant under both infinitesimal gauge transformations and infinitesimal tetrad rotations, since \(\Psi_0\) and \(\Psi_1\) vanish on the background, so no gauge or tetrad conversion is required. What must be matched are the coupling expansion and the projected radial, angular and Kinnersley-tetrad normalizations of the cited ingoing-radiation-gauge equation.
\end{enumerate}

\subsection{Background geometry, Weyl scalar and perturbations}
\label{sec:background_geometry}

We begin by specifying the background geometry. As stated above, we restrict attention to nonrotating BH solutions. In dCS gravity, the nonrotating BH background coincides with the Schwarzschild metric, given in Eq.~\eqref{eq:schw}~\cite{Jackiw:2003pm,Grumiller:2007rv,Yunes:2009hc}. In the nonrotating limit, the background scalar field vanishes, so that the scalar sector enters only through the perturbation. We perturb the metric and scalar field using Eqs.~\eqref{eq:perts} and~\eqref{eq:sfperts}
and decompose the perturbations into tensor and scalar spherical harmonics using Eqs.~\eqref{eq:hodd} and \eqref{eq:heven}. 
For each $(\ell,m)$ mode and QNM frequency $\omega$, the odd- and even-parity metric perturbations in GR can be encoded in the gauge-invariant RW and ZM master functions introduced previously. 
In dCS gravity, however, the odd-parity gravitational perturbation couples to the scalar perturbation whereas the even-parity sector remains decoupled in the nonrotating limit and is described independently by the master variable $\Psi_{\rm ZM}(r)$. 

To construct the curvature-based description, we next fix a Newman-Penrose null tetrad, $l^\mu, n^\mu, m^\mu$, and $\bar m^\mu$, on the Schwarzschild background. In coordinates $x^{\mu} = (t,r,\theta,\phi)$, 
we take 
\begin{subequations}\label{eq:bg_tetrad}
\begin{align}
l^\mu &= \left(\frac{r}{r-2M},\,1,\,0,\,0\right),
\label{eq:bg_tetrad_l}\\
n^\mu &= \left(\frac{1}{2},\, -\frac{1}{2}+\frac{M}{r},\,0,\,0\right),
\label{eq:bg_tetrad_n}\\
m^\mu &= \left(0,\,0,\,\frac{1}{\sqrt{2}\,r},\,\frac{i}{\sqrt{2}\,r}\csc\theta\right),
\label{eq:bg_tetrad_m}\\
\bar m^\mu &=\left(0,\,0,\,\frac{1}{\sqrt{2}\,r},\,-\frac{i}{\sqrt{2}\,r}\csc\theta\right),
\label{eq:bg_tetrad_mbar}
\end{align}
\end{subequations}
which satisfy the standard inner-product relations
\begin{equation}
l_{\mu} n^{\mu}=-1,
\qquad
m_{\mu} \bar m^{\mu}=1,
\label{eq:bg_tetrad_inner}
\end{equation}
with all other contractions vanishing. This is the tetrad used in the Weyl contraction defining $\Psi_0$ in the nonrotating limit.

The Newman-Penrose Weyl scalar relevant for the Chandrasekhar transformation in our sign convention is
\begin{equation}
\Psi_0:=C_{\mu \nu \rho \sigma}\,l^{\mu} m^{\nu} l^{\rho} m^{\sigma} \,,
\label{eq:psi0_def}
\end{equation}
where $C_{\mu \nu \rho \sigma}$ is the Weyl tensor. Perturbing the metric as in Eq.~\eqref{eq:perts}, we define $\delta\Psi_0$ to be the part of $\Psi_0$ linear in $h_{\mu\nu}$. The goal is then to express $\delta\Psi_0$ in terms of the RW, ZM, and scalar master variables, thereby relating the curvature-based and metric-based descriptions of the dCS perturbation problem.

\subsection{Parity decomposition and harmonic separation}
\label{sec:parity_decomp}

As discussed in Sec.~\ref{sec:metpert}, the metric perturbation can be decomposed into even- and odd-parity sectors, using the standard tensor-harmonic decomposition introduced there. 
Accordingly, we write
\begin{equation}
h_{\mu\nu}=h^{\rm (even)}_{\mu\nu}+h^{\rm (odd)}_{\mu\nu}.
\label{eq:parity_metric_split}
\end{equation}
Substituting this decomposition into the tetrad contraction in Eq.~\eqref{eq:psi0_def}, and retaining only terms linear in the perturbations, yields
\begin{equation}
\delta\Psi_0
=
\delta\Psi_0^{\rm (even)}
+
\delta\Psi_0^{\rm (odd)},
\label{eq:psi0_parity_split}
\end{equation}
where the separation into parity sectors is exact at linear order. In particular, before harmonic decomposition, the even- and odd-parity contributions can already be identified directly from the explicit expression for $\delta\Psi_0$.

We then decompose each sector into tensor spherical harmonics and Fourier modes. For each $(\ell,m)$ multipole, we allow the even- and odd-parity channels to carry, in principle, different complex frequencies $\omega_E$ and $\omega_O$, respectively. For the gravitational-led odd branch one may set $\omega_O=\omega_{O,g}$, but the local reconstruction formulas themselves only require a fixed odd-sector Fourier frequency. The separated form may therefore be written as
\begin{widetext}
\begin{equation}
\delta\Psi_0(t,r,\theta,\phi)
=
A_{\rm even}^{\ell m}(\theta,\phi)\,
e^{-i\omega_E t}\,
\delta\Psi_{0,\ell m}^{\rm (even)}(r;\omega_E)
+
A_{\rm odd}^{\ell m}(\theta,\phi)\,
e^{-i\omega_O t}\,
\delta\Psi_{0,\ell m}^{\rm (odd)}(r;\omega_O),
\label{eq:psi0_sep}
\end{equation}
\end{widetext}
where $A_{\rm even}^{\ell m}$ and $A_{\rm odd}^{\ell m}$ denote the residual angular structures that remain after simplification. These quantities are not physical amplitudes, but rather angular factors associated with the chosen harmonic basis and tetrad projection. For a nonrotating BH in dCS gravity, these functions are given by
\begin{subequations}  \label{eq:Alabels}
\begin{align}
    A_{\text{odd}} &= -2i\big(-\partial_\theta^2 + \cot\theta\,\partial_\theta + \csc^2\theta\,\partial_\phi^2\big)Y_{\ell m} \nonumber \\  &\qquad - 4\csc\theta\big(\partial_\theta\partial_\phi - \cot\theta\,\partial_\phi\big)Y_{\ell m} \,, \nonumber \\
    &= 2i\sqrt{(\ell-1)\ell(\ell+1)(\ell+2)}\; {}_{+2}Y_{\ell m} \,, \\
    A_{\text{even}} &= \frac{i}{2}A_{\text{odd}} \,,
\end{align}   
\end{subequations}
where ${}_{+2}Y_{\ell m}$ are spin-weighted spherical harmonic. Using the orthogonality properties of spin-weighted spherical harmonics, one can then eliminate the angular dependence to obtain a transformation in $(t,r)$ coordinates only, if required.
In the nonrotating problem, the QNM frequencies are \(m\)-degenerate, but the \(m\) label is nevertheless retained because the waveform amplitudes and angular harmonics carry it. The curvature and tensor-harmonic reconstruction formulas below are intended for \(\ell\ge2\). Scalar monopole and dipole perturbations are allowed in the scalar sector, but they do not enter these spin-\(\pm2\) curvature and tensor-radiative reconstruction maps.

For the construction of the Chandrasekhar transformation, the essential point is that the radial even- and odd-parity pieces can be treated independently at this stage. One may therefore rewrite $\delta\Psi_{0,\ell m}^{\rm (even)}$ in terms of the polar master variable $\Psi_{\rm ZM}$ and its radial derivative, and $\delta\Psi_{0,\ell m}^{\rm (odd)}$ in terms of the axial master variable $\Psi_{\rm RW}$, its radial derivative, and, in dCS gravity, the scalar mode. Equation~\eqref{eq:psi0_sep} is precisely the separated structure used later in assembling the full nonrotating Chandrasekhar transformation.

\subsection{Odd-parity channel, dCS coupling, and the RW master function}
\label{sec:oddchandra}

We first focus on the odd-parity contribution to the perturbed Weyl scalar appearing in Eq.~\eqref{eq:psi0_sep}. Prior to rewriting it in terms of master variables, this contribution can be expressed schematically in terms of the axial metric perturbations as
\begin{align}
\delta\Psi_{0,\ell m}^{\rm (odd)}(r)
&=
A_{\rm pref}(r)\Bigl[
A_1(r)\,h_0(r)+A_2(r)\,h_1(r) \nonumber \\
& \qquad +A_3(r)\,h_0'(r)+A_4(r)\,h_1'(r)
\Bigr],
\label{eq:psi0odd_Adecomp}
\end{align}
where
\begin{subequations}\label{eq:Acoeffs_def}
\begin{align}
A_{\rm pref}(r)
&=\frac{1}{4r^2(r-2M)^2},
\label{eq:Apref_def}
\\
A_1(r)&=-(2M+i r^2\omega_{O}),
\label{eq:A1_def}
\\
A_2(r)&= i r\omega_{O}(2M-r),
\label{eq:A2_def}
\\
A_3(r)&= -r(2M-r),
\label{eq:A3_def}
\\
A_4(r)&= (2M-r)^2.
\label{eq:A4_def}
\end{align}
\end{subequations}
The coefficients \(A_i(r)\) are determined entirely by the chosen tetrad and the Schwarzschild background geometry, and are algebraic functions of \(r\), \(M\), and the generic odd-sector QNM frequency \(\omega_O\).

The next step is to rewrite the axial metric perturbations in terms of the RW master function. For nonzero QNM frequency, \(\omega_O\neq0\), as appropriate for the ringdown modes considered here, the standard axial reconstruction relations give
\begin{align}
h_{0}(r)
&=
\frac{i(2M-r)}{r^{3}\omega_{O}}
\left[
-2M\,h_{1}(r)
+
(2M-r) r h_{1}'(r)
\right],
\label{eq:h0_from_h1}
\\[4pt]
h_{1}(r)
&=
\frac{r}{f(r)}
\Psi_{\rm RW}(r).
\label{eq:h1_from_RW}
\end{align}
Substituting these relations into Eq.~\eqref{eq:psi0odd_Adecomp}, the odd-parity curvature perturbation can be written as a linear combination of $\Psi_{\rm RW}$ and its first two radial derivatives,
\begin{align}
\delta\Psi_{0,\ell m}^{\rm (odd)}(r)
&=
B_{\rm pref}(r)\Bigl[
B_1(r)\,\Psi_{\rm RW}(r)+B_2(r)\,\frac{d\Psi_{\rm RW}}{dr}(r) \nonumber \\
& \qquad +B_3(r)\,\frac{d^2\Psi_{\rm RW}}{dr^2}(r)
\Bigr],
\label{eq:psi0odd_Bdecomp}
\end{align}
where
\begin{subequations}\label{eq:Bcoeffs_def}
\begin{align}
B_{\rm pref}(r)
&=\frac{i}{4\,r^3(r-2M)^2\,\omega_{O}},
\label{eq:Bpref_def}
\\
B_1(r)
&=
r^2\left[
2i\omega_O(3M-r)-r^2\omega_O^2
\right],
\label{eq:B1_odd_correct}
\\
B_2(r)
&=
2r(2M-r)
\left[
2M-r+ir^2\omega_O
\right],
\label{eq:B2_odd_correct}
\\
B_3(r)
&=
r^2(2M-r)^2
.
\label{eq:B3_odd_correct}
\end{align}
\end{subequations}
The coefficients $B_i(r)$ are again algebraic functions of $r$, $M$, and $\omega_{O}$.

At this stage, the expression in Eq.~\eqref{eq:psi0odd_Bdecomp} is not yet in Chandrasekhar form, since it still contains a second radial derivative of the master function. In GR, one would use the RW equation to eliminate this term. In dCS gravity, the corresponding expression is given by Eq.~\eqref{eq:RW} and can be rewritten as 
\begin{align}
\label{eq:dCSRWnorot}
\Psi_{\rm RW}''(r)
&=
\frac{
\mathcal{Q}_{\ell}(r)\Psi_{\rm RW}(r)
+
2M(2M-r)r\Psi_{\rm RW}'(r)
}{
r^2(r-2M)^2
}
\nonumber \\
&-
\frac{
96\pi iM(2M-r)\alpha\omega_O
}{
r^4(r-2M)^2
}
R_{\ell m}(r),
\end{align}
where
\begin{align}
\mathcal{Q}_{\ell}(r) &= 12M^2-2(\Lambda+3)Mr+r^2(\Lambda-r^2\omega_O^2)\, , 
\label{eq:Qell_RW_definition}
\end{align}
where \(\Lambda\) is defined in Eq
The above axial master equation contains a scalar-induced source term, and this is precisely how the scalar perturbation enters the gravitational curvature map when using Eq.~\eqref{eq:dCSRWnorot} to eliminate the second derivatives of the master function, 
\begin{widetext}
\begin{align}
\label{eq:chandra_odd_final}
\delta\Psi_{0,\ell m}^{\rm odd}(r;\omega_O)
&=
\frac{i}{4r^3(r-2M)^2\omega_O}
\left[
\widetilde B_1(r)\Psi_{\rm RW}(r)
+
\widetilde B_2(r)\Psi_{\rm RW}'(r)
\right]
+
\frac{24\pi M\alpha}{r^5(2M-r)}
R_{\ell m}(r)\,,
\end{align}
\end{widetext}
where
\begin{align}
\widetilde B_1(r)
&=
12M^2
+
2Mr\left(
-3-\ell-\ell^2+3ir\omega_O
\right)
\nonumber \\
&+
r^2\left[
\ell+\ell^2
-
2r\omega_O(i+r\omega_O)
\right].
\label{eq:B1tilde_odd_correct_expanded}
\\
\widetilde B_2(r)
&=
2(2M-r)r
\left[
3M+r(-1+ir\omega_O)
\right].
\label{eq:B2tilde_odd_correct}
\end{align}
The above equation reduces exactly to the Chandrasekhar map in~\cite{Chandrasekhar:1975nkd} with $\Psi_{\RW} \to \Psi_{\RW}^{\GR}$ when one sets $\alpha = 0$. Notice that the dCS modification is explicitly proportional to $\alpha$, and not $\alpha^2$, because $R_{\ell m}$ is sourced by $\alpha R^\star R$ (see Eq.~\eqref{eq:CSfield}), so the last term is of $\mathcal{O}(\alpha^2)$.

Equation~\eqref{eq:chandra_odd_final} shows explicitly that, in the axial channel, the curvature perturbation $\delta\Psi_0$ acquires a term directly proportional to the scalar perturbation $R_{\ell m}$. This contribution arises unambiguously from the sequence of steps used in the construction:
\begin{enumerate}[label=(\roman*)]
    \item computing $\delta\Psi_0$ from the perturbed Weyl tensor in the fixed tetrad;
    \item expressing the axial metric perturbations in terms of $\Psi_{\rm RW}(r)$;
    \item using the dCS-modified axial RW equation to eliminate the second radial derivative.
\end{enumerate}
Thus, the scalar dependence of the Chandrasekhar map is a direct consequence of the axial gravitational-scalar coupling in the linearized dCS field equations.

\subsection{Even-parity channel and the ZM master function}
\label{sec:evenchandra}

We now follow a procedure analogous to that of Sec.~\ref{sec:oddchandra} for the even-parity contribution $\delta\Psi_{0,\ell m}^{\rm (even)}(r;\omega_E)$ appearing in Eq.~\eqref{eq:psi0_sep}. After extracting the even-parity angular dependence and specializing to the nonrotating tetrad, the radial even-parity contribution reduces to
\begin{align}
\delta\Psi_{0,\ell m}^{\rm (even)}(r;\omega_E)
&=\frac{1}{4r^{2}(r-2M)^{2}}
\Bigl[
r^{2}H_{0}(r)
\nonumber\\
&+(2M-r)\bigl(-2r\,H_{1}(r)+(2M-r)\,H_{2}(r)\bigr)
\Bigr],
\label{eq:psi0_even_H}
\end{align}
which serves as the starting point for rewriting the curvature perturbation in terms of a gauge-invariant polar master variable.

We next express the polar metric perturbations appearing in Eq.~\eqref{eq:psi0_even_H} in terms of the ZM master function $\Psi_{\rm ZM}(r)$. Operationally, this reduction proceeds by substituting the polar reconstruction relations for the metric perturbations $H_2$ and $H_0$, and are given by~\cite{Zerilli:1971wd,Moncrief:1974am,Chandrasekhar:1975nkd,Maggiore:2018sht,Chandrasekhar_1983}
\begin{align} \label{eq:H0dcs}
H_2(r) &= \frac{1}{f(r)^2}H_0(r) \,, \\ 
H_{0}(r)&=\mathcal{H}_{1}(r)\,H_{1}(r)+\mathcal{K}(r)\,K(r),
\end{align}
where
\begin{subequations}
\begin{align}
\mathcal{H}_{1}(r)&=\frac{i f(r)\bigl(2r^{3}\omega_{E}^{2}-M\Lambda\bigr)}{r (6M+\lambda r) \omega_{E}},
\\
\mathcal{K}(r)&=\frac{\lambda r^{2}f(r)+2M(r-3M)-2r^{4}\omega_{E}^{2}}{r (6M+\lambda r)}\,,
\end{align}
\end{subequations}
with $\Lambda$ and \(\lambda\) given by Eq.~\eqref{eq:lambda_def}. Finally we redefine the remaining metric perturbations in terms of the ZM master function as
\begin{align}
H_1(r) &= -2i\omega_{E} \left[\frac{\lambda r(r-3M)-6M^{2}}{(r-2M)\bigl(6M+\lambda r\bigr)}\, \Psi_{\rm ZM}(r)
+
r\,\Psi_{\rm ZM}'(r)\right] \,, \\
\label{eq:Kdcs}
K(r) &= \frac{24M^{2}+6\lambda M r+\lambda\Lambda r^{2}}{r^{2}\bigl(6M+\lambda r\bigr)}\,\Psi_{\rm ZM}(r)
+
2f(r)\,\Psi_{\rm ZM}'(r)\,.
\end{align}
Any higher than first-order derivatives can be reduced by using the ZM equation given in Eq.~\eqref{eq:ZM}.

Substituting Eqs.~\eqref{eq:H0dcs}-\eqref{eq:Kdcs} into Eq.~\eqref{eq:psi0_even_H}, the final result is therefore a closed expression of the form
\begin{align}
\delta\Psi_{0,\ell m}^{\rm (even)}(r;\omega_E)
=& \,
C_{\rm even}(r;\ell,\omega_E)\,\Psi_{\rm ZM}(r) \nonumber \\
& \qquad+
D_{\rm even}(r;\ell,\omega_E)\, \Psi_{\rm ZM}'(r)
\label{eq:chandra_even_struct}
\end{align}
where
\begin{subequations}\label{eq:CevenDeven_def}
\begin{align}
C_{\rm even}(r;\ell,\omega_E)
&=
-\frac{N_1(r;\ell,\omega_E)}
{2r^3(r-2M)^2\bigl(6M+\lambda r\bigr)^2}
\nonumber\\
&\quad
+\frac{N_2(r;\ell,\omega_E)}
{r^2(r-2M)^2\bigl(6M+\lambda r\bigr)^2},
\label{eq:Ceven_def}
\\[0.5em]
D_{\rm even}(r;\ell,\omega_E)
&=
\frac{\lambda r(r-3M)-6M^{2}}
     {r^{2}(r-2M)\bigl(6M+\lambda r\bigr)}
-\frac{i\omega_E}{r-2M},
\label{eq:Deven_def}
\end{align}
\end{subequations}
with
\begin{subequations}\label{eq:Neven_def}
\begin{align}
N_1(r;\ell,\omega_E)
&=
\Bigl(24M^{2}+6\lambda M r+\lambda\Lambda r^{2}\Bigr)
\nonumber\\
&\quad \times
\Bigl(2r^{4}\omega_E^{2}-\lambda r^{2}f(r)-2M(r-3M)\Bigr),
\label{eq:N1even_def}
\\[0.5em]
N_2(r;\ell,\omega_E)
&=
\Bigl(6M^{2}-\lambda r(r-3M)\Bigr)
\nonumber\\
&\quad \times
\Bigl(\Lambda M+i r\omega_E\bigl(6M+\lambda r\bigr)-2r^{3}\omega_E^{2}\Bigr),
\label{eq:N2even_def}
\end{align}
\end{subequations}
where $\Lambda$ and \(\lambda\) given by Eq.~\eqref{eq:lambda_def}
The above expression differs by a factor of two from the common Zerilli convention \(\lambda_Z=(\ell-1)(\ell+2)/2\). All formulas in this section use the convention in Eq.~\eqref{eq:lambda_def}.

Equation~\eqref{eq:chandra_even_struct} gives the even-parity contribution to the Chandrasekhar transformation in terms of the polar master variable. In contrast to the odd-parity sector discussed in Sec.~\ref{sec:oddchandra}, no scalar contribution appears here in the nonrotating limit of dCS gravity. This is consistent with the structure of the perturbation equations reviewed earlier, in which the scalar field couples only to the axial gravitational sector, while the polar sector remains decoupled. Consequently, the even-parity curvature map is fully determined by the ZM master function and its first radial derivative. In dCS gravity, at the nonrotating order, since the polar metric perturbations do not gain any additional deviations, the corresponding Chandrasekhar transformation given in Eq.~\eqref{eq:chandra_even_struct} is exactly as in GR.

\subsection{Full nonrotating transformation}
\label{sec:full_chandra}

Combining the odd- and even-parity results obtained in Secs.~\ref{sec:oddchandra} and \ref{sec:evenchandra}, we now have all ingredients for the full Chandrasekhar transformation in dCS gravity. For convenience, we rewrite Eq.~\eqref{eq:psi0_sep} below:
\begin{align}
\delta\Psi_{0,\ell m}(t,r,\theta,\phi)
&=
A_{\rm even}^{\ell m}(\theta,\phi)\,e^{-i\omega_E t}\,
\delta\Psi_{0,\ell m}^{\rm (even)}(r;\omega_E)
\nonumber \\ 
&+
A_{\rm odd}^{\ell m}(\theta,\phi)\,e^{-i\omega_{O} t}\,
\delta\Psi_{0,\ell m}^{\rm (odd)}(r;\omega_{O}).
\label{eq:chandra_full}
\end{align}
Here, $A_{\rm even}$ and $A_{\rm odd}$ are given by Eqs.~\eqref{eq:Alabels}, and $\delta\Psi_{0,\ell m}^{\rm (even)}$ and $\delta\Psi_{0,\ell m}^{\rm (odd)}$ are given by Eqs.~\eqref{eq:chandra_even_struct} and \eqref{eq:chandra_odd_final}, respectively. Equation~\eqref{eq:chandra_full} is the final nonrotating Chandrasekhar transformation for dCS gravity, written in terms of the polar and axial metric master variables together with the scalar perturbation. Here \(\Psi_{\rm RW}\) and \(R_{\ell m}\) solve the coupled odd-sector equations, Eqs.~\eqref{eq:SF} and \eqref{eq:RW}, at the \textit{same} QNM frequency \(\omega_O\). For a quasinormal mode, \(\Psi_{\rm RW}\) and \(R_{\ell m}\) are two components of a single coupled eigenvector, and \(\omega_O\) labels whichever of the two pole families (scalar-led or gravity-led) is under consideration.

A corresponding transformation following this procedure can be constructed for the Weyl scalar $\Psi_4$ and is presented in Appendix~\ref{app:psi4_chandra}. In GR, \(\Psi_0\) and \(\Psi_4\) are related by the Teukolsky-Starobinsky identities~\cite{Starobinsky:1973aij,Starobinsky:1974nkd,Press:1973zz,Teukolsky:1974yv}. We use this only as background motivation. Schematically, in a fixed radial normalization one may write
\begin{subequations}\label{eq:TS_identity}
\begin{align}
{\cal T}^{\rm GR}_{4\to0}[\Psi_4^{\rm GR}]
&= C_{\ell\omega}^{\rm GR}\,\Psi_0^{\rm GR},
\\
{\cal T}^{\rm GR}_{0\to4}[\Psi_0^{\rm GR}]
&= \widetilde C_{\ell\omega}^{\rm GR}\,\Psi_4^{\rm GR}.
\end{align}
\end{subequations}
Here \({\cal T}^{\rm GR}_{4\to0}\) and \({\cal T}^{\rm GR}_{0\to4}\) denote the usual fourth-order GR radial operators of the Teukolsky-Starobinsky identities, while the constants depend on the normalization of the tetrad and Teukolsky radial functions. Since these operators are not used in the dCS reconstruction below, we do not show an explicit operator normalization here.

At present, however, we \emph{do not} assume that the Teukolsky-Starobinsky identities remain unchanged in dCS gravity. Since the axial gravitational sector is modified through its coupling to the scalar perturbation, the relation between $\Psi_0$ and $\Psi_4$ may also be modified, or may require a reformulation in terms of an enlarged coupled or sourced system. Instead, we present a consistency check between the transformations obtained for the $\Psi_0$ and $\Psi_4$ Weyl scalar perturbations in Appendix~\ref{app:psi0_psi4_reduced_check}.

\subsection{Consistency check: substitution into the modified Teukolsky equation}
\label{sec:mte_check}

The Chandrasekhar transformation in Eq.~\eqref{eq:chandra_full} is useful only
if the curvature perturbation reconstructed from the metric master variables
satisfies the curvature-based evolution equation of the modified Teukolsky
formalism. We therefore substitute Eq.~\eqref{eq:chandra_full} into the
nonrotating limit of the modified Teukolsky equation in the ingoing radiation
gauge~\cite{Wagle:2023fwl,Li:2025fci}, and reduce the result using the master
equations derived above: the polar equation for $\Psi_{\rm ZM}$ in the
even-parity sector, and the dCS-modified axial equation
in Eq.~\eqref{eq:dCSRWnorot} in the odd-parity sector. 

Before the two descriptions can be compared, one normalization must be made
explicit. The modified Teukolsky formalism of~\cite{Wagle:2023fwl,Li:2025fci}
organizes the perturbative expansion in the dimensionless coupling
\begin{equation}
\zeta \equiv \frac{\alpha^{2}}{\kappa M^{4}} \,,
\label{eq:zeta_def}
\end{equation}
and absorbs a factor of $\zeta^{1/2}$ into the scalar field, so that the
quantity $\vartheta^{(1,1)}$ appearing in the source of the modified Teukolsky
equation is \emph{not} the physical scalar perturbation but the physical field
with one half power of the coupling removed. In the metric-perturbation
language used here, the scalar radial function $R_{\ell m}$ is the physical
field, and the two are related by
\begin{equation}
\Theta_{\ell m} \;=\; \zeta^{-1/2}\,R_{\ell m}\,,
\qquad
\zeta^{1/2} = \frac{\alpha}{\sqrt{\kappa}\,M^{2}} \,.
\label{eq:theta_Rlm}
\end{equation}
This relation cannot be absorbed into an overall rescaling. Because
$\delta\Psi_0^{\rm dCS}$ is quadratic in the coupling while the source of the
modified Teukolsky equation is linear in $\Theta_{\ell m}$, the \emph{half}
power in Eq.~\eqref{eq:theta_Rlm} makes the gravitational and scalar sectors
rescale by different powers. We therefore expand in $\zeta$, as
in Eq.~\eqref{eq:zeta_def}, in order to match the perturbative coefficients of
Refs.~\cite{Wagle:2023fwl,Li:2025fci}. Throughout this check
we use the GR-connected hierarchy
\begin{equation}
\Psi_{\rm RW}=\Psi_{\rm RW}^{(0)}+\zeta\,\Psi_{\rm RW}^{(1)}+\cdots,
\qquad
R_{\ell m}=\zeta^{1/2}\Theta_{\ell m}+\cdots,
\label{eq:gr_connected_hierarchy}
\end{equation}
in which the explicit $\alpha R_{\ell m}$ term of the curvature map contributes
at $\mathcal{O}(\zeta)$. The check does not probe the alternative power
counting associated with independently prescribed $\mathcal{O}(1)$ scalar data.

With this identification the check is passed exactly. In the even-parity
sector, a single use of the ZM equation reduces the substituted expression
to zero identically in $(r,M,\omega_E,\ell)$: the polar Chandrasekhar map
carries ZM solutions into homogeneous solutions of the Teukolsky
equation, consistent with the absence of any source in the even sector at
$\mathcal{O}(\zeta^{1},\chi^{0})$. In the odd-parity sector, all dependence on
the axial master function $\Psi_{\rm RW}$ and its derivatives cancels
identically, and the remainder reproduces the scalar-driven source of the
modified Teukolsky equation term by term, with no residual and no free
constant. At $\mathcal{O}(\zeta^{0})$ both sectors reduce to the classical
Chandrasekhar identity of GR~\cite{Chandrasekhar:1975nkd,Chandrasekhar_1983},
which provides an independent check of the coefficient functions
$\widetilde{B}_{1,2}$, $C_{\rm even}$ and $D_{\rm even}$.

We emphasize that this is a nontrivial verification rather than a
restatement. The scalar-dependent term in Eq.~\eqref{eq:chandra_odd_final} is
not free: it is fixed by the derivation, and the substitution tests both its
functional form and its coefficient against the modified Teukolsky equation
derived in Refs.~\cite{Wagle:2023fwl,Li:2025fci}, whose operator
and scalar source are imported here rather than rederived.
The algebra underlying this section, including the explicit reduction rules and the separate even- and odd-parity residuals, is provided in the supplementary notebook~\cite{Pratikgit}.

\subsection{Remarks on usage and interpretation}
\label{sec:remarks_usage}

In this section, we present brief remarks on the relevance of these results and compare these with some recent works.

\vspace{0.2cm}
\noindent \textit{a.~Frequency separation and isospectrality breaking.}

Allowing distinct complex frequencies $\omega_E$ and $\omega_{O}$ in Eq.~\eqref{eq:chandra_full} is essential once isospectrality is broken. In GR, the odd- and even-parity sectors of a Schwarzschild BH are isospectral, and a single complex QNM frequency can be associated with both sectors for a chosen multipole. In dCS gravity, by contrast, the odd-parity gravitational perturbation couples to the scalar field, while the even-parity sector remains decoupled in the nonrotating limit. The two sectors therefore need not share the same oscillation frequency or damping rate. The present construction treats them as independent frequency-domain channels, with the angular factors $A_{\rm even}^{\ell m}$ and $A_{\rm odd}^{\ell m}$ carrying the residual harmonic information.

\vspace{0.2cm}
\noindent \textit{b.~Gauge-invariant content.}

Although $\Psi_0$ depends on the choice of null tetrad, the construction presented here is performed in a fixed background tetrad and expresses $\delta\Psi_0$ entirely in terms of gauge-invariant master variables, namely $\Psi_{\rm ZM}$, $\Psi_{\rm RW}$, and the scalar radial mode. This is the natural notion of a Chandrasekhar map in the present setting: a map between the curvature representation and the gauge-invariant metric master-system representation, written in a fixed background tetrad. As noted below, the residual frame dependence of \(\delta\Psi_0\) is only a spin-boost weight, not a general tetrad dependence.

\vspace{0.2cm}
\noindent \textit{c.~Practical role.}

Equation~\eqref{eq:chandra_full} provides a direct route for extracting the gauge-invariant curvature variable
\(\delta\Psi_0\) once the RW/ZM master variables and the scalar perturbation are known. Conversely, it also serves as a stringent algebraic consistency check on the modified Teukolsky equation itself. From a practical standpoint, the transformation therefore bridges two complementary descriptions of the same perturbation problem: the short-range RW/ZM system, which is often more convenient for QNM and scattering calculations, and the Teukolsky-type curvature formalism, which is natural for radiation-gauge reconstruction and curvature-based observables. Because \(\delta\Psi_0\) and \(\delta\Psi_4\) vanish on any Petrov type-D background, like the Schwarzschild metric considered here, they are gauge invariant by the Stewart-Walker lemma~\cite{Stewart:1974uz,Teukolsky:1973ha,Wald:1973zaa} and are defined up to spin-boost weight in the Kinnersley tetrad. The displayed coefficients are exterior-domain expressions and are not horizon regular at \(r=2M\)~\cite{Kinnersley:1969zza,Chatziioannou:2012rf,Geroch:1973am,Hawking:1972hy}.

\vspace{0.2cm}
\noindent \textit{d.~Curvature-level scalar dependence versus tensor-strain reconstruction.}

Equation~\eqref{eq:chandra_odd_final} shows that the odd-parity Chandrasekhar map contains an explicit term proportional to the scalar perturbation \(R_{\ell m}(r)\). This is a curvature-level statement: the scalar degree of freedom enters \(\delta\Psi_0\) through the coupled axial-scalar dCS dynamics. It should not, however, be interpreted as a direct scalar contribution to the detector tensor strain. The tensor polarizations are obtained only after an asymptotic metric reconstruction at infinity. As shown later in Sec.~\ref{sec:dcsobs}, the leading far-zone expressions for \(h_+\) and \(h_\times\) contain no independent local algebraic term proportional to \(R_{\ell m}\). They are reconstructed only from the gravitational radiative master functions.

This distinction is consistent with the general frequency-contamination mechanism discussed in Ref.~\cite{Lestingi:2025jyb} and with the waveform-level ringdown amplitude framework of Ref.~\cite{Crescimbeni:2025kxi}, but the detailed phenomenology is not the focus of the present paper. For the purposes of this work, the relevant conclusion is narrower. In nonrotating dCS gravity, the scalar field appears explicitly in the odd-parity curvature map, while the leading tensor-strain reconstruction contains no separate local scalar term. Any scalar-frequency contribution to \(h_+\) or \(h_\times\) must therefore enter indirectly through an induced radiative odd metric response, after the appropriate RW-to-CPM normalization, rather than by reading off the scalar term in \(\delta\Psi_0\). The residue-level and phenomenological consequences of such scalar-frequency tensor responses are left to a companion analysis~\cite{paper2}.

\section{Observables in GR}
\label{sec:grobs}

As discussed in Sec.~\ref{sec:review}, the gauge-invariant master variables introduced earlier are directly related to the physically observable gravitational radiation. In particular, in the asymptotic region far from the source, they determine the waveform measured by a distant observer and the associated energy flux. In this section, we review the relation between the odd- and even-parity master functions and the observable polarization amplitudes $h_+$ and $h_\times$ in the transverse-traceless (TT) gauge. Our discussion follows the standard treatment of Schwarzschild perturbations in the wave zone~\cite{Cunningham:1978zfa,Cunningham:1979px,Nagar:2005ea}.

To extract the radiative content of the perturbation, we project the metric perturbation onto the orthonormal tetrad of stationary observers in the Schwarzschild background. Denoting tetrad indices by hatted symbols, the projected perturbation is
\begin{equation}
h_{\hat{\mu}\hat{\nu}}
=
e^{\mu}{}_{\hat{\mu}}\,e^{\nu}{}_{\hat{\nu}}\,h_{\mu\nu},
\label{eq:h_tetrad_proj}
\end{equation}
with the tetrad being the diagonal matrix
\begin{equation}
e^{\mu}{}_{\hat{\mu}}
=
\textrm{diag}\left\{
f(r)^{-1/2},\, f(r)^{1/2},\, r^{-1},\, (r\sin\theta)^{-1}
\right\},
\label{eq:obs_tetrad}
\end{equation}
where recall that $f(r)=1-2M/r$. In the wave zone, the perturbation is required to satisfy the outgoing-radiation conditions
\begin{align}
\left\{
h_{\hat{\theta}\hat{\theta}},
h_{\hat{\phi}\hat{\phi}},
h_{\hat{\theta}\hat{\phi}}
\right\}
&=
\mathcal{O}(r^{-1}),
\nonumber \\
h_{\hat{\mu}\hat{\nu}}
&=
\mathcal{O}(r^{-2})
\quad
\text{for all other components},
\label{eq:wave_zone_scaling}
\end{align}
together with the leading-order tracelessness condition
\begin{equation}
h_{\hat{\theta}\hat{\theta}}+h_{\hat{\phi}\hat{\phi}}
=
\mathcal{O}(r^{-2}).
\label{eq:tracefree_condition}
\end{equation}
These conditions ensure that the asymptotic perturbation coincides with the usual TT waveform measured at infinity. For the remainder of this section, we work in an asymptotic radiation gauge in which the leading \(\mathcal{O}(r^{-1})\) metric perturbation is transverse and traceless. We do not assume that a global TT gauge exists at finite radius.

\subsection{Odd-parity contribution}
\label{sec:odd_obs_gr}

We first consider odd-parity perturbations. Using the decomposition in
Eq.~\eqref{eq:hodd} and stripping the axial harmonics, the leading tetrad components in the radiation zone
satisfy
\begin{align}
h^{(o)}_{\hat{\theta}\hat{t}}
&=
\frac{h^{(o)}_0}{r}
\sim \mathcal{O}(r^{-2}),
\label{eq:hodd_thetat}
\\
h^{(o)}_{\hat{\theta}\hat{r}}
&=
\frac{h^{(o)}_1}{r}
\sim \mathcal{O}(r^{-2}),
\label{eq:hodd_thetar}
\end{align}
so $h_0^{(o)}=\mathcal{O}(r^{-1}) = h_1^{(o)}$ asymptotically. 
The corresponding odd-parity contributions to the two GW polarizations are
\begin{align}
h^{(o)}_+
&\equiv
\frac{1}{2}
\left(
h^{(o)}_{\hat{\theta}\hat{\theta}}
-
h^{(o)}_{\hat{\phi}\hat{\phi}}
\right)
\nonumber\\
&=
\frac{h}{r^2}
\left(
\nabla_\theta S_\theta
-
\frac{\nabla_\phi S_\phi}{\sin^2\theta}
\right)
+\mathcal{O}(r^{-2}),
\label{eq:hodd_plus}
\\
h^{(o)}_\times
&\equiv
h^{(o)}_{\hat{\theta}\hat{\phi}}
=
\frac{2h}{r^2}
\frac{\nabla_{(\phi}S_{\theta)}}{\sin\theta}
+\mathcal{O}(r^{-2}).
\label{eq:hodd_cross}
\end{align}
Since the $\hat{\theta}\hat{\theta}$ and $\hat{\theta}\hat{\phi}$ angular components of the hatted metric perturbations must fall off as $r^{-1}$ as discussed in Eq.~\eqref{eq:wave_zone_scaling}, the odd-parity metric amplitude $h$ in Eq.~\eqref{eq:hodd} is of $\mathcal{O}(r)$, and so $h_{+,\times}^{(o)} = \mathcal{O}(1/r)$. 

We now relate the angular metric perturbation entering
Eqs.~\eqref{eq:hodd_plus} and \eqref{eq:hodd_cross} to the radiative
odd-parity master function. A notational distinction is important here.
The object that directly reconstructs the odd-parity contribution to the
asymptotic strain is the Cunningham-Price-Moncrief (CPM)-normalized odd master function. We denote it by~\cite{Cunningham:1978zfa,Cunningham:1979px,Nagar:2005ea,Martel:2005ir}
\begin{equation}
\Psi^{\rm rad}_{O,\ell m}
\equiv
\Psi_{\rm CPM,\ell m}.
\label{eq:psi_rad_cpm_def}
\end{equation}
The CPM function satisfies the same homogeneous RW operator as Eq.~\eqref{eq:RW_freq_domain}, up to the normalization convention relating it to the field-equation variable. It should therefore not be confused with an arbitrary RW-normalized variable used to write the axial field equation. If the latter variable is denoted by \(\Psi_{\rm RW,\ell m}\), then the two are related by an exact time-domain normalization map. 
\begin{equation}
\Psi_{{\rm RW},\ell m}
=
-\partial_t \Psi_{\rm CPM,\ell m},
\label{eq:rw_cpm_time_dictionary}
\end{equation}
This is the vacuum relation between the field-equation and radiative odd normalizations. A generic source modifies it by an explicit source projection as shown in the dCS case in Sec.~\ref{sec:dcsobs_odd}. The frequency-domain form below is therefore applied mode by mode, to the outgoing residue at each pole,
\begin{equation}
\Psi^{\rm rad}_{O,\ell m}
\equiv
\Psi_{\rm CPM,\ell m}
=
-\frac{i}{\omega_{O}}\,
\Psi_{\rm RW,\ell m}.
\label{eq:rw_cpm_freq_dictionary}
\end{equation}
All subsequent formulas that use an exact RW-to-radiative transfer factor assume this convention. If a different RW normalization is used in the field equation, the outgoing residue must be converted to \(\Psi^{\rm rad}_{O}\) before it is inserted into the strain.

Using the \(\theta\phi\) component of the perturbed field equations, the angular metric
perturbation is related to the radiative odd master function by
\begin{equation}
\partial_t h
=
f(r)\,\partial_r\!\bigl(r\Psi^{\rm rad}_{O}\bigr)
+
h_0^{(o)}.
\label{eq:h_radodd_relation}
\end{equation}
This relation is the key step in the asymptotic waveform calculation,
because it links the metric amplitude \(h\), which enters \(h_+\) and
\(h_\times\), to the CPM-normalized odd master function. At large \(r\),
outgoing-wave behavior gives
\[
f(r)\partial_r\!\bigl(r\Psi^{\rm rad}_{O}\bigr)
\simeq
-\partial_t\!\bigl(r\Psi^{\rm rad}_{O}\bigr),
\]
while \(h_0^{(o)}=\mathcal{O}(r^{-1})\). 
Hence,
Eq.~\eqref{eq:h_radodd_relation} implies
\begin{equation}
\frac{h}{r}
=
-\Psi^{\rm rad}_{O}
+
\mathcal{O}(r^{-1})
=
-\Psi_{\rm CPM}
+
\mathcal{O}(r^{-1}).
\label{eq:h_over_r_cpm}
\end{equation}

Substituting Eq.~\eqref{eq:h_over_r_cpm} into
Eqs.~\eqref{eq:hodd_plus} and \eqref{eq:hodd_cross} gives
\begin{align}
h^{(o)}_+
&=
-\frac{\Psi^{\rm rad}_{O}}{r}
\left(
\nabla_\theta S_\theta
-
\frac{\nabla_\phi S_\phi}{\sin^2\theta}
\right)
+\mathcal{O}(r^{-2}),
\label{eq:hodd_plus_cpm}
\\
h^{(o)}_\times
&=
-\frac{2\Psi^{\rm rad}_{O}}{r}
\frac{\nabla_{(\phi}S_{\theta)}}{\sin\theta}
+\mathcal{O}(r^{-2}).
\label{eq:hodd_cross_cpm}
\end{align}
Using the standard relation between the axial angular structures and the
spin-weight \(-2\) spherical harmonics, the odd-parity contribution can be
written compactly as
\begin{align}
\bigl(h^{(o)}_+ - i h^{(o)}_\times\bigr)_{\ell m}
=&
\frac{i}{r}
\sqrt{\frac{(\ell+2)!}{(\ell-2)!}}\,
\Psi^{\rm rad}_{O,\ell m}\,
{}_{-2}Y_{\ell m}(\theta,\phi)
\nonumber\\
&\qquad +
\mathcal{O}(r^{-2})
\nonumber\\
=&
\frac{i}{r}
\sqrt{\frac{(\ell+2)!}{(\ell-2)!}}\,
\Psi_{{\rm CPM},\ell m}\,
{}_{-2}Y_{\ell m}(\theta,\phi)
+
\mathcal{O}(r^{-2}).
\label{eq:hodd_compact}
\end{align}

Equivalently, if the axial perturbation problem has been solved in terms of an RW-normalized field-equation variable \(\Psi_{\rm RW}\), then the same expression may be written mode by mode as
\begin{align}
\bigl(h^{(o)}_+ - i h^{(o)}_\times\bigr)_{\ell m }
=&
\frac{1}{r \omega_O}
\sqrt{\frac{(\ell+2)!}{(\ell-2)!}}\,
\Psi_{{\rm RW},\ell m}\,
\nonumber\\
&\qquad
\times {}_{-2}Y_{\ell m}(\theta,\phi)
+ \mathcal{O}(r^{-2}),
\label{eq:hodd_compact_rw_dictionary}
\end{align}
Thus, the RW and CPM variables have the same QNM pole locations, but their
residues differ by the normalization factor as shown in Eq.~\eqref{eq:rw_cpm_freq_dictionary}. This distinction is irrelevant
for identifying the frequencies, but it is relevant for waveform
amplitudes and for any residue-level statement about which modes enter the
tensor strain. If one instead uses a literature convention in which the field-equation axial variable differs by an overall factor, this renormalization changes by the inverse factor, and all waveform formulas below therefore use \(\Psi^{\rm rad}_{O}\) as the invariant radiative normalization.

\subsection{Even-parity contribution}
\label{sec:even_obs_gr}

We now consider the even-parity sector. Using Eq.~\eqref{eq:heven} and projecting onto the orthonormal tetrad, the two GW polarizations are
\begin{align}
h^{(e)}_+
&\equiv
\frac{1}{2}
\left(
h^{(e)}_{\hat{\theta}\hat{\theta}}
-
h^{(e)}_{\hat{\phi}\hat{\phi}}
\right)
\nonumber\\
&=
\frac{G}{2}
\left(
\nabla_\theta\nabla_\theta Y_{\ell m}
-
\frac{\nabla_\phi\nabla_\phi Y_{\ell m}}{\sin^2\theta}
\right)
=
\frac{G}{2}\,W_{\ell m},
\label{eq:heven_plus}
\\
h^{(e)}_\times
&\equiv
h^{(e)}_{\hat{\theta}\hat{\phi}}
=
G\,\frac{\nabla_\theta\nabla_\phi Y_{\ell m}}{\sin\theta}
=
\frac{G}{2}\,\frac{X_{\ell m}}{\sin\theta},
\label{eq:heven_cross}
\end{align}
where $W_{\ell m}$ and $X_{\ell m}$ are the standard even-parity angular functions,
\begin{align}
W_{\ell m}
&=\left(\nabla_\theta\nabla_\theta-
\frac{\nabla_\phi\nabla_\phi}{\sin^2\theta}
\right)Y_{\ell m},
\\
X_{\ell m}
&=2\nabla_\theta\nabla_\phi Y_{\ell m}.
\end{align}
With the axial orientation fixed above, the spin-weighted identities used are
\begin{align}
W_{\ell m}-i\frac{X_{\ell m}}{\sin\theta}
&=
\sqrt{\frac{(\ell+2)!}{(\ell-2)!}}\,{}_{-2}Y_{\ell m},
\\
\left(\nabla_\theta S_\theta-
\frac{\nabla_\phi S_\phi}{\sin^2\theta}\right)
-i\frac{2\nabla_{(\phi}S_{\theta)}}{\sin\theta}
&=
-i\sqrt{\frac{(\ell+2)!}{(\ell-2)!}}\,{}_{-2}Y_{\ell m}.
\end{align}

In the wave zone, the remaining even-parity amplitudes satisfy
\begin{equation}
H_2=\mathcal{O}(r^{-2}),
\qquad
h_1^{(e)}=\mathcal{O}(r^{-1}),
\label{eq:even_wavezone_scaling}
\end{equation}
so they do not contribute at leading order to the radiative waveform. The tracelessness condition in Eq.~\eqref{eq:tracefree_condition} then implies
\begin{equation}
K
=
\frac{\Lambda}{2}\,G
+
\mathcal{O}(r^{-2}),
\label{eq:K_asymp_G}
\end{equation}
where $\Lambda=\ell(\ell+1)$.

Using the gauge-invariant combinations introduced earlier in Eqs.~\eqref{eq:k1}-\eqref{eq:q1} and writing
\(E\equiv K-({\Lambda}/{2})G\) for the deviation from
Eq.~\eqref{eq:K_asymp_G}, one obtains the exact identity
\begin{align}
q_1
&=
\frac{\Lambda}{2}\left[(\Lambda-2)r+6M\right]G
+\left[(\Lambda-2)r+6M\right]E
\nonumber\\
&\quad
-2r^{2}f(r)\,\partial_r E
-2\Lambda f(r)\,h^{(e)}_{1}
+2rf(r)^{2}H_{2}.
\label{eq:q1_exact}
\end{align}
The \(r^{2}\partial_r G\) terms cancel between \(r\Lambda k_1\) and
\(4rfk_2\) only once the explicit factors of \(f\) in \(k_1\) and the
connection term \(MK/(2rf)\) in \(k_2\) are both retained, and the surviving
\(3\Lambda MG\) is precisely what produces the \(6M\) of the Zerilli
denominator. Substituting Eq.~\eqref{eq:q1_exact} into the definition of the
ZM master function gives
\begin{align}
\Psi_{\rm ZM}
&=
\frac{rq_1}{\Lambda\left[r(\Lambda-2)+6M\right]} \nonumber \\
&=
\frac{1}{2}rG
+\frac{rE}{\Lambda}
+\frac{2rf\left(rfH_{2}-\Lambda h^{(e)}_{1}-r^{2}\partial_r E\right)}
{\Lambda\left[(\Lambda-2)r+6M\right]} .
\label{eq:psi_even_exact}
\end{align}
For an outgoing vacuum solution in an asymptotically TT gauge, the last group
does not contribute at leading order, and one recovers the standard asymptotic
reconstruction~\cite{Moncrief:1974am,Nagar:2005ea}
\begin{equation}
\Psi_{\rm ZM}
=
\frac{1}{2}rG
+
\mathcal{O}(r^{-1}).
\label{eq:psi_even_asymp}
\end{equation}
Establishing this limit from Eq.~\eqref{eq:K_asymp_G} alone would require solving the asymptotic vacuum constraints one order further, since \(E=\mathcal{O}(r^{-2})\) does not by itself bound \(r^{2}f\,\partial_r E\)~\cite{Nagar:2005ea}.
Hence,
\begin{equation}
G
=
\frac{2\Psi_{\rm ZM}}{r}
+
\mathcal{O}(r^{-2}).
\label{eq:G_from_psi_even}
\end{equation}
Substituting Eq.~\eqref{eq:G_from_psi_even} into Eqs.~\eqref{eq:heven_plus} and \eqref{eq:heven_cross}, and again using the standard angular identities, one obtains
\begin{equation}
\bigl(h^{(e)}_+ - i h^{(e)}_\times\bigr)_{\ell m}
=
\frac{1}{r}
\sqrt{\frac{(\ell+2)!}{(\ell-2)!}}\,
\Psi_{{\rm ZM},\ell m}\,
{}_{-2}Y_{\ell m}(\theta,\phi)
+
\mathcal{O}(r^{-2}).
\label{eq:heven_compact}
\end{equation}

\subsection{Combined asymptotic waveform}
\label{sec:combined_obs_gr}

The full asymptotic waveform is obtained by summing the even- and odd-parity contributions,
\begin{equation}
h_+
=
h^{(e)}_+ + h^{(o)}_+,
\qquad
h_\times
=
h^{(e)}_\times + h^{(o)}_\times.
\label{eq:h_total_def}
\end{equation}
Using Eqs.~\eqref{eq:hodd_compact} and \eqref{eq:heven_compact}, the complex GW strain may be written as
\begin{align}
h_+ - i h_\times
&=
\frac{1}{r}
\sum_{\ell,m}
\sqrt{\frac{(\ell+2)!}{(\ell-2)!}}
\left(
\Psi_{{\rm ZM},\ell m}
+
i\,\Psi^{\rm rad}_{O,\ell m}
\right)
\nonumber\\
&\qquad
\times {}_{-2}Y_{\ell m}(\theta,\phi) +\mathcal{O}(r^{-2}).
\label{eq:h_complex_full}
\end{align}
This expression makes explicit that, in GR, the full radiative content of the perturbation is encoded in the even ZM function \(\Psi_{{\rm ZM},\ell m}\) and the odd CPM-normalized radiative function \(\Psi^{\rm rad}_{O,\ell m}\). If the odd dynamics is solved using \(\Psi_{{\rm RW},\ell m}\), Eq.~\eqref{eq:hodd_compact_rw_dictionary} supplies the required residue conversion.

It is also common to express the odd-parity contribution in terms of the Moncrief or RW field-equation master function \(Q^{(o)}\). In the convention of Refs.~\cite{Cunningham:1978zfa,Cunningham:1979px,Nagar:2005ea}, it is related asymptotically to the CPM/radiative function by
\begin{equation}
\Psi^{\rm rad}_{O}(t)
=
-\int_{-\infty}^{t} Q^{(o)}(t')\,dt'.
\label{eq:Psiodd_Qodd_relation}
\end{equation}
The integration constant is fixed by requiring \(h/r\to0\) at early times for asymptotically flat data, and is set to zero throughout [cf.\ Eq.~(70) of Ref.~\cite{Nagar:2005ea}].
In that case, Eq.~\eqref{eq:h_complex_full} becomes
\begin{align}
h_+ - i h_\times
&=
\frac{1}{r}
\sum_{\ell,m}
\sqrt{\frac{(\ell+2)!}{(\ell-2)!}}
\left[
\Psi_{{\rm ZM},\ell m}
-
i\int_{-\infty}^{t}Q^{(o)}_{\ell m}(t')\,dt'
\right]
\nonumber\\
&\qquad \times
{}_{-2}Y_{\ell m}(\theta,\phi)
+\mathcal{O}(r^{-2}).
\label{eq:h_complex_Qodd}
\end{align}

The forms in Eqs.~\eqref{eq:h_complex_full} and \eqref{eq:h_complex_Qodd} are mathematically equivalent, but it is important to distinguish their normalization. The CPM/radiative function \(\Psi^{\rm rad}_{O}\) has dimensions of length and contributes directly to the waveform amplitude, whereas \(Q^{(o)}\) is dimensionless in this convention and enters the waveform only after time integration. This distinction is important when comparing conventions in the literature, and in particular when computing the asymptotic strain.

\section{Observables in dynamical Chern-Simons gravity}
\label{sec:dcsobs}

In this section, we extend the discussion of Sec.~\ref{sec:grobs} to dCS gravity. As reviewed in Sec.~\ref{sec:RWZM-bGR}, the theory contains a nonminimal coupling between the metric and a dynamical scalar field. At the level of linear perturbations, this coupling modifies the structure of the master equations. In particular, the even-parity gravitational sector remains decoupled, whereas the odd-parity gravitational sector couples directly to the scalar perturbation through the system of Eqs.~\eqref{eq:SF}--\eqref{eq:ZM}. As a result, the Schwarzschild isospectrality of GR is broken: the even-parity gravitational branch and the odd-parity gravitational-led branch are no longer characterized by the same QNM spectrum.

A central physical question is then whether the scalar perturbation affects the observable gravitational-wave polarizations. Put differently, once the odd-parity gravitational perturbation is coupled to the scalar field at the level of the master equations, does this coupling leave a direct imprint on the asymptotic waveform, or does it enter only through the gravitational master functions? Answering this question is essential for understanding how isospectrality breaking manifests itself at the level of gravitational-wave measurements.

The question has two logically distinct parts. The first is local and kinematical: does the leading far-zone reconstruction of \(h_+\) and \(h_\times\) contain a separate algebraic term proportional to the scalar radial function \(R_{\ell m}\)? The second is spectral and perturbative: can a homogeneous scalar QNM source an \(\mathcal{O}(\zeta)\) field-equation RW response, for GR-connected ringdown data and away from near-degeneracy, whose CPM-normalized radiative odd projection appears in the tensor strain at the scalar frequency? The calculation in this section answers the first question. The second requires a statement about the pole content, RW-to-CPM tensor projection, and excitation of the coupled odd-sector response, and is deferred to the companion phenomenology paper~\cite{paper2}.

To address the reconstruction question, we repeat the asymptotic waveform analysis of Sec.~\ref{sec:grobs} in the context of dCS gravity. The strategy is the same: we work in the wave zone, impose the outgoing-radiation and leading-order tracelessness conditions, and relate the asymptotic metric perturbations to the gauge-invariant master variables. 
The even-parity sector is straightforward. Since the polar master equation, Eq.~\eqref{eq:ZM}, is independent of the scalar perturbation in the nonrotating theory, the derivation of the even-parity waveform proceeds exactly as in GR. Therefore, the polar contribution to the observable strain remains unchanged and is still given by Eq.~\eqref{eq:heven_compact}. In particular, the scalar field does not enter the even-parity gravitational-wave polarizations at this order.

The odd-parity sector is qualitatively different. There, the axial master equation, Eq.~\eqref{eq:RW}, is coupled to the scalar master equation, Eq.~\eqref{eq:SF}. This opens the possibility that the scalar perturbation could modify the odd-parity contribution to the waveform. The purpose of the following analysis is to determine whether such scalar dependence survives as a direct local term in the asymptotic gravitational-wave observables. As we show below, despite the nontrivial coupling in the axial field equations and the associated breaking of isospectrality, the observable plus and cross polarizations remain free of direct scalar reconstruction terms in the nonrotating dCS case. 
For GR-connected merger data, in which the scalar is itself generated at
\(\mathcal{O}(\alpha)\), this does not exclude an \(\mathcal{O}(\zeta) = \mathcal{O}(\alpha^2) \) scalar-frequency tensor response through the odd gravitational channel; it establishes that any such term must enter exclusively through the radiative odd metric master function, after the RW-to-CPM conversion if necessary, rather than through an independent local scalar term in the strain.

\subsection{Odd parity}
\label{sec:dcsobs_odd}

In this subsection, we determine whether the scalar perturbation contributes directly to the odd-parity gravitational-wave polarizations. This can be checked by deriving the far-zone polarization formulas from the metric perturbations and then using the dCS perturbed field equations in the nonrotating limit. The relation
between the GW polarizations \(h_+\) and \(h_\times\) and the metric
perturbations is purely kinematical, and is therefore independent of the
underlying theory of gravity. Hence, the expressions derived in
Sec.~\ref{sec:odd_obs_gr} (especially Eqs.~\eqref{eq:hodd_cross} and ~\eqref{eq:hodd_plus}) remain valid in dCS gravity. 

As in the GR case, the relevant equation is the \(\theta\phi\) component
of the perturbed field equations. This equation connects the metric
perturbation \(h\), which enters the TT waveform, to the odd-parity master
variable. In nonrotating dCS gravity, before imposing the RW gauge, the
\(\theta\phi\) perturbed field equation takes the form
\begin{widetext}
\begin{align}
2 i r^4 \omega_O\, h_0(r)
&+8 M^2 r^2 h_1'(r)-8 M r^3 h_1'(r)+2 r^4 h_1'(r)
+4 M r(r-2M) h_1(r)
\nonumber\\
&-4 M^2 r^2 h_2''(r)+4 M r^3 h_2''(r)-r^4 h_2''(r)
+12 M^2 r h_2'(r)-10 M r^2 h_2'(r)+2 r^3 h_2'(r)
\nonumber\\
&-\left(16 M^2-12Mr+r^4\omega_O^2+2r^2\right)h_2(r)
=0 .
\label{eq:34eqdcsNR}
\end{align}
\end{widetext}
Here \(\omega_O\) denotes the frequency of the odd parity component being reconstructed. For a gravitational-led branch, \(\omega_O=\omega_{O,g}\). Notice that Eq.~\eqref{eq:34eqdcsNR} is already independent of the scalar field perturbation due to the nature of the C-tensor given in Eq.~\eqref{eq:CTensor} and it is discussed it further detail in Appendix~\ref{app:ctensor_thetaphi}. For an induced metric response at a scalar-led frequency, \(\omega\) should instead be evaluated at the corresponding scalar-led pole. In the above equation, we have already used the harmonic decomposition to factor out the angular dependence\footnote{ Equivalently, one may express the angular dependence using the Legendre polynomials, after expanding the vector spherical harmonics defined in Eq.~\eqref{eq:hodd}.}. The displayed equation is used only as the projected field equation needed for the leading large-radius reconstruction below; no finite-radius master-equation or spectral statement is inferred from this component alone. Importantly, this expression is obtained without imposing the RW gauge, as in GR.

We now introduce the RW-normalized odd-parity master variable using the same gauge-invariant definition adopted earlier,
\begin{equation}
\Psi_{{\rm RW},\ell m}(r)
=
\frac{f(r)}{r}
\left[
h_1(r)
-
\frac{r^2}{2}\,
\partial_r\!\left(\frac{h_2(r)}{r^2}\right)
\right].
\label{eq:RWnorotdcs_obs}
\end{equation}
This is not the variable used in the axial field equation given in Eq.~\eqref{eq:RW} as the Eq.~\eqref{eq:RW} is written in the RW gauge whereas Eq.~\eqref{eq:RWnorotdcs_obs} is a general form outside the RW gauge. It should also be distinguished from the CPM-normalized radiative odd master function, which we denote in the same way as in GR in Eq.~\eqref{eq:psi_rad_cpm_def}.  

For nonzero QNM frequency, \(\omega\neq0\), the vacuum relation between the 
RW and the CPM function in Eq.~\eqref{eq:rw_cpm_time_dictionary} acquires an explicit local correction in the sourced dCS problem. 
Writing \(\mathcal{L}=\partial_{r_*}^2-\partial_t^2-V^{A}_{\rm eff}\), Eq.~\eqref{eq:RW} reads \(\mathcal{L}\Psi_{{\rm RW} \ell m} =-\left(6\alpha M f/\kappa r^{5}\right)\partial_t R_{\ell m}\), and direct evaluation of the sourced axial constraint gives
\begin{align}
\Psi_{{\rm RW},\ell m}(t,r)
=
-\partial_t\!\left[
\Psi_{{\rm CPM},\ell m}(t,r)
-
\Delta(r)\,R_{\ell m}(t,r)
\right]\,,
\label{eq:rw_cpm_time_dictionary_sourced}
\end{align}
where
\begin{equation}
    \Delta(r)=\frac{6\alpha M}{\kappa\,\lambda \,r^{3}}\,,
\end{equation}
with \(\lambda\) given in Eq.~\eqref{eq:lambda_def}.
Equation~\eqref{eq:rw_cpm_time_dictionary_sourced} holds at every value of the radial coordinate. The gravitational-wave observables, however, are defined at null infinity, so for the strain reconstruction we need only its \(r\to\infty\) form, in which the shift due to the scalar field perturbation drops out.
Since \(\Delta=\mathcal{O}(r^{-3})\) while \(\Psi_{\rm CPM}=\mathcal{O}(1)\) for an outgoing mode, the shift contributes to \(\partial_t h\) at \(\mathcal{O}(r^{-2})\) against an \(\mathcal{O}(r)\) leading term. The leading null-infinity strain and the per-pole residue relation below are therefore unaffected. For the normalization in Eq.~\eqref{eq:RWnorotdcs_obs}, the two are related asymptotically, mode by mode, through the same expression as in GR, given in Eq.~\eqref{eq:rw_cpm_freq_dictionary}.
This normalization factor does not change the QNM pole locations, but it does change the residues and therefore the observable amplitudes.

Solving Eq.~\eqref{eq:RWnorotdcs_obs} for \(h_1(r)\) and substituting the
result into Eq.~\eqref{eq:34eqdcsNR}, the projected \(\theta\phi\) equation reduces to
\begin{align}
&2 i \omega_O\, h_0(r)
-
\omega_O^2 h_2(r)
\nonumber \\
&\qquad -
\frac{2(2M-r)}{r}
\left[
r \partial_r\Psi_{{\rm RW},\ell m}(r)
+
\Psi_{{\rm RW},\ell m}(r)
\right]
=
0 .
\label{eq:34eqdcsNRsim}
\end{align}
Equation~\eqref{eq:34eqdcsNRsim} is valid without imposing a gauge condition. To identify the angular metric amplitude \(h\) with the radiative strain, however, we now specialize to an asymptotic radiation gauge in which the leading \(\mathcal{O}(r^{-1})\) metric perturbation is transverse and traceless. In this gauge, \(h_0=\mathcal{O}(r^{-1})\) and \(h_1=\mathcal{O}(r^{-1})\), while \(h=\mathcal{O}(r)\). Taking the large-radius outgoing limit at a fixed pole,
\[ 
\partial_r\Psi_{{\rm RW},\ell mn} \simeq i\omega_{\ell mn}\Psi_{{\rm RW},\ell mn},
\]
Eq.~\eqref{eq:34eqdcsNRsim} then gives
\begin{equation}
\frac{h}{r}
=
\frac{i}{\omega_{\ell mn}}\,
\Psi_{{\rm RW},\ell mn}
+
\mathcal{O}(r^{-1})
=
-\Psi^{\rm rad}_{O,\ell mn}
+
\mathcal{O}(r^{-1}) .
\label{eq:dcs_h_over_r_cpm}
\end{equation}
The factor in Eq.~\eqref{eq:dcs_h_over_r_cpm} follows from the convention \(h_2=2h\). Equivalently, Eq.~\eqref{eq:34eqdcsNRsim} first gives \(h_2/r=2i\Psi_{{\rm RW},\ell m n}/\omega_{\ell m n}\), and only then \(h/r=i\Psi_{{\rm RW},\ell m n}/\omega_{\ell m n}\). The right-hand equality, written in terms of \(\Psi^{\rm rad}_{O}\), is pole-independent and is the form used in the strain formulas below.

Equation~\eqref{eq:34eqdcsNRsim} is the \(\theta\phi\) field equation evaluated at the perturbative order used throughout this work, namely through quadratic order in the dCS coupling \(\alpha\). The scalar-dependent contribution to this component vanishes identically at finite radius, not merely asymptotically: as shown in Appendix~\ref{app:ctensor_thetaphi}, the linearized C-tensor about Schwarzschild with vanishing background scalar has \(\delta C_{\theta\phi}[\delta\vartheta]=0\), and \(\delta T^{\vartheta}_{\mu\nu}=0\) at linear order; the scalar coupling enters instead through the axial master equation. Thus, no term proportional to the scalar radial function \(R_{\ell m}\) appears in this component, making the scalar-free local reconstruction a consequence of the evaluated field equation, not an additional assumption.
Equation~\eqref{eq:34eqdcsNRsim} has the same asymptotic reconstruction structure as the corresponding equation in GR. This is the key result. Although the axial master equation is coupled to the scalar perturbation, the local field equation that reconstructs the leading tensor polarizations does not contain the scalar radial function as an independent algebraic contribution. The scalar can affect the result only through the solution for the odd-parity metric master variable and through the corresponding RW-to-CPM normalization.

Substituting Eq.~\eqref{eq:dcs_h_over_r_cpm} into Eqs.~\eqref{eq:hodd_plus} and \eqref{eq:hodd_cross}, and using the standard relation between the axial angular structures and the spin-weight \(-2\) spherical harmonics, the odd-parity contribution to the tensor strain is given by the same expression as GR, presented in Eq.~\eqref{eq:hodd_compact}. However, the radiative wavefunction \(\Psi^{\rm rad}_{O,\ell m}(r)\) in dCS gravity now includes the GR-order wavefunction as well as a modification that would be introduced due to dCS gravity. 
Equivalently, using Eq.~\eqref{eq:rw_cpm_freq_dictionary}, the outgoing residue at a given pole may be written in terms of the RW-normalized variable as
\begin{align}
\bigl(h^{(o)}_+ - i h^{(o)}_\times\bigr)_{\ell m}
=&
\frac{1}{\omega_{O}\, r}
\sqrt{\frac{(\ell+2)!}{(\ell-2)!}}\,
\Psi_{{\rm RW},\ell m}(r)\,
\nonumber\\
&
\times {}_{-2}Y_{\ell m}(\theta,\phi)
+
\mathcal{O}(r^{-2}) .
\label{eq:obsodd_rw_variable}
\end{align}
Equation~\eqref{eq:obsodd_rw_variable} is a per-pole statement, valid for the outgoing residue at \(\omega_{O}\). It is not a time-domain operation on an arbitrary sourced superposition. Because the coupled odd sector has both gravitational-led and scalar-led pole families, a full time-domain waveform should be assembled either directly from the CPM radiative variable using Eq.~\eqref{eq:obsodd_rw_variable}, which requires no frequency factor, or by converting the RW Green-function residues pole by pole with Eq.~\eqref{eq:rw_cpm_freq_dictionary} before summing.

Equation~\eqref{eq:obsodd_rw_variable} shows explicitly that scalar perturbations make no direct local contribution to the \(+\) and \(\times\) polarizations. Combined with the results of Ref.~\cite{Wagle:2019mdq}, this implies that scalar perturbations contribute no direct local term to any GW polarization. The same conclusion follows from
\begin{equation}
    \delta\psi_4 \propto \partial_t^2\left(h_+ - i h_\times\right)\,,
\end{equation}
together with the results of Sec.~\ref{sec:oddchandra} and Appendix~\ref{app:psi4_chandra}. On direct substitution of \(\delta\psi_4\) it may appear that the scalar field perturbation contributes to \(h_+\) and \(h_\times\). However, these observables are defined at null infinity, and there the scalar contribution decays faster than that of the gravitational sector. The two main results of this work therefore complement each other. The absence of an explicit scalar contribution should not, however, be read as the absence of scalar-led modes in the gravitational channel, which arise from the coupled structure of the perturbed field equations, Eq.~\eqref{eqs:dcsperturbed_nonrot}. The effect of scalar-led modes on \(h_+\) and \(h_\times\) is discussed in the companion paper~\cite{paper2}.

\subsection{Even parity}
\label{sec:dcsobs_even}

We now turn to the even-parity sector. In dCS gravity, the even-parity sector remains completely decoupled, as seen from Eq.~\eqref{eq:ZM}. Consequently, there is no direct scalar contribution to the polar gravitational-wave polarizations or any additional modifications due to the bGR theory. In the nonrotating case, the even-parity master equation has exactly the same structure as in GR at the order considered here, and therefore the derivation of the asymptotic waveform proceeds identically to the analysis presented in Sec.~\ref{sec:even_obs_gr}. As a result, the even-parity contribution to the gravitational-wave strain is unchanged and is given by Eq.~\eqref{eq:heven_compact}.
Thus, in the nonrotating dCS case, the polar tensor observables are identical in form to those of GR.

\subsection{Scalar-frequency tensor responses}
\label{sec:scalar_frequency_brief}

The reconstruction result above should not be read as a statement that scalar-led frequencies can never appear in the tensor waveform. It states only that the leading local far-zone map contains no independent algebraic term proportional to \(R_{\ell m}\). A scalar-led pole of the coupled odd system can still contribute to \(h_+\) and \(h_\times\) if that pole has a nonzero radiative odd metric residue and is excited by the source. A purely particular response to a monochromatic gravitational-led source carries the driving gravitational frequency. Scalar-led frequencies enter through the homogeneous or Green-function pole part of the scalar response, with coefficients fixed by the projection of the full metric-scalar ringdown data onto the coupled pole. 
For GR-connected ringdown data, the projection of the full metric-scalar data onto a scalar-led pole is generically \(\mathcal{O}(\alpha)\), while the metric component of the scalar-led eigenmode is also \(\mathcal{O}(\alpha)\). The corresponding tensor-strain contribution therefore begins at \(\mathcal{O}(\alpha^2)=\mathcal{O}(\zeta)\), away from a near-degeneracy. This projection can be generated by the metric-sector forcing, or equivalently by the preceding retarded source history, even if the scalar data happen to vanish on a chosen slice. 
Schematically,
\begin{equation}
h^{\infty}_{\ell m,{\rm sf}}(u)
=
\zeta
\sum_k
\widehat{C}^{\,\infty}_{O,s,\ell mk}
 e^{-i\omega^{s,0}_{\ell k}u}
+
O(\zeta^2).
\label{eq:scalar_frequency_tensor_response_short}
\end{equation}
Here \(u=t-r_*\), and \(h^{\infty}_{\ell m,{\rm sf}}\) denotes the coefficient of the \(1/r\) complex tensor strain with the spin-weighted angular factor stripped. The expansion uses the unshifted scalar frequency and is a fixed-ringdown-time expansion and it is not uniform for \(u\sim1/\zeta\). \(\widehat{C}^{\,\infty}_{O,s,\ell mk}\) includes the excitation and the induced CPM-normalized tensor projection. This is not a direct scalar polarization and is not obtained by reading off the scalar term in \(\delta\Psi_0\).

This interpretation is consistent with the frequency-contamination picture of Ref.~\cite{Lestingi:2025jyb} and with the waveform-level ringdown amplitude framework of Ref.~\cite{Crescimbeni:2025kxi}, where ringdown models include both shifted gravitational-led modes and extra-field-induced modes. The additional content of the present calculation is the local reconstruction statement specific to nonrotating dCS gravity: any extra-field-induced mode that appears in \(h_+\) or \(h_\times\) must enter through the coupled odd metric response, after the appropriate RW-to-CPM radiative normalization. The beating, effective-frequency behavior, and data-analysis impact of these scalar-frequency tensor responses are deferred to the companion paper~\cite{paper2}.

\section{Conclusions}
\label{sec:conclusion}

In this work, we have developed the relation between curvature perturbations, metric master variables, and detector tensor polarizations for nonrotating BHs in dCS gravity, extending the classical Schwarzschild Chandrasekhar transformation~\cite{Chandrasekhar:1975nkd,Chandrasekhar_1983} to a theory in which the odd-parity gravitational sector is coupled to a dynamical scalar field. The central result is a separation between two statements that are otherwise easily conflated. The odd-parity Chandrasekhar map for the perturbed Weyl scalar $\delta \Psi_0^{\rm odd}$ in terms of the RW master function $\Psi_{\RW}$ and its derivative $\Psi_{\RW}'$ contains an explicit scalar perturbation ($R_{\ell m})$ dependent term, reflecting the coupled nature of the axial gravitational-scalar system. On the other hand, the leading tensor polarizations \(h_+\) and \(h_\times\) contain no independent local algebraic term proportional to the scalar radial function.

This separation rests on two independent pieces of the calculation. At the curvature level, the even-parity contribution to the Weyl scalar perturbations are expressed through the ZM function and retain the GR structure at the order considered. The odd-parity contribution, on the other hand, is expressed through the RW function and the scalar radial mode, with the scalar term arising when the dCS-modified axial equation is used to eliminate higher radial derivatives. We checked that our calculated Chandrasekhar map is consistent through the substitution of this map into the modified Teukolsky equation. We further showed that the newly-derived Chandrasekhar map satisfies the modified Teukolsky equation computed in~\cite{Wagle:2023fwl}, once both are normalized to the same basis. 
At the waveform level and for the GW polarizations \(h_+\) and \(h_\times\), the even-sector strain is reconstructed from \(\Psi_{\rm ZM}\), while the odd-sector strain is reconstructed from the CPM-normalized radiative variable \(\Psi^{\rm rad}_O\equiv\Psi_{\rm CPM}\), which is related to the RW master function. We also show that no term proportional to \(R_{\ell m}\) survives in the GW polarizations because the field equations describing the radiative metric perturbations are independent of the scalar field perturbations. Moreover, asymptotically, the scalar field perturbation falls off at a faster rate than the GW wavefunction.

These results imply that explicit scalar dependence in a curvature variable is not, by itself, evidence for a direct scalar term in the measured polarizations. In nonrotating dCS gravity, the scalar field can affect the tensor waveform only through the gravitational metric response: a scalar-led pole of the coupled odd system contributes to \(h_+\) and \(h_\times\) only if it carries a nonzero radiative odd metric residue that is excited.  
For the GR-connected ringdown problem considered here, such a contribution requires a nonzero radiative odd metric residue and a nonzero projection of the full coupled metric-scalar data inherited from the preceding inspiral-merger evolution, or equivalently of the retarded source history, onto the scalar-led pole. That projection is generically $\mathcal{O}(\alpha)$, and the metric component of the scalar-led eigenmode is also $\mathcal{O}(\alpha)$, so the amplitude of the tensor response begins at $\mathcal{O}(\alpha^2)=\mathcal{O}(\zeta)$. 
Such a contribution is therefore a scalar-frequency tensor response, conceptually distinct from an independent scalar polarization or from a local scalar term read directly off the Weyl scalar perturbations. This work therefore makes clear the general frequency-contamination picture proposed for beyond-GR ringdowns in~\cite{Lestingi:2025jyb,Crescimbeni:2024qca,Crescimbeni:2025kxi}: the present calculation identifies the specific, theory-level channel through which such contamination can or cannot occur in nonrotating dCS gravity.

Several extensions follow directly from our work. The companion paper will use these reconstruction results to analyze parity-split beating, scalar-frequency tensor residues, and effective-frequency diagnostics for realistic ringdown signals~\cite{paper2}. Extending the curvature-to-metric dictionary constructed here to rotating dCS BHs is the most pressing open problem: in the rotating case, the RW/ZM decomposition is no longer available, so the corresponding reconstruction will likely require metric-reconstruction methods developed in \cite{Cohen_Kegeles_1975, Kegeles:1979an, Chrzanowski:1975wv, Keidl:2010pm, Green:2019nam, Toomani:2021jlo, Hollands:2024iqp, Aksteiner:2016mol, Wardell:2024yoi, Loutrel:2020wbw, Ripley:2020xby, Li:2026rkf} with proper extensions to beyond-GR theories. The same strategy, a curvature map first and an asymptotic strain reconstruction second, is not specific to dCS gravity and could be applied to other parity-violating or higher-curvature theories with a similar sector-coupling structure, including higher-derivative gravity~\cite{Cano:2023tmv} and Einstein-dilaton-Gauss-Bonnet gravity~\cite{Blazquez-Salcedo:2016enn,Chung:2024ira,Chung:2024vaf}. Comparing the perturbative reconstruction hierarchy derived here against existing numerical-relativity simulations of dCS mergers~\cite{Okounkova:2019dfo,Okounkova:2019zjf} would also test whether the leading-order absence of an independent local scalar term in \(h_+\) and \(h_\times\) persists into the nonlinear, strong-field regime.

Ringdown tests of the Kerr hypothesis are now entering the precision regime. For GW250114, the loudest GW event observed to date, the dominant quadrupolar frequency is constrained to within a few percent of the GR prediction using the full inspiral-merger-ringdown signal, while direct spectroscopy of the dominant mode and its first overtone probes the Kerr QNM spectrum at the tens-of-percent level~\cite{LIGOScientific:2025wao}. Ringdown and remnant  consistency tests have also been extended to the GWTC-4 population, including  new events from O4a together with events from previous observing runs~\cite{LIGOScientific:2026wpt}. Independently, theory-agnostic searches for additional field-induced modes in the ringdown are being actively developed~\cite{Crescimbeni:2024qca,Crescimbeni:2025kxi}.
Any such search, whether theory-specific or theory-agnostic, implicitly relies on a mapping between the bulk content of a modified-gravity perturbation problem and the modes that can appear in the strain. The results presented here make that mapping explicit for the simplest isospectrality-breaking setting, and provide the reconstruction dictionary needed to correctly attribute, or correctly exclude, a scalar-frequency signature as next-generation ground-based detectors~\cite{Reitze:2019iox} and space-based interferometers~\cite{Baker:2019nia} push ringdown measurements to still higher precision.

\section{Acknowledgements}

We thank Colin Weller, Hector Silva, Luis Lehner and Alessandra Buonanno for interesting discussions. P.~W. acknowledges funding from the Deutsche Forschungsgemeinschaft (DFG) project number: 386119226. D.~L. and N.~Y. acknowledge support from the Simons Foundation (via Award No. 896696), the Simons Foundation International (via Grant No. SFI-MPS-BH-00012593-01), and the NSF (via Grants No.~PHY-2512423). Y.~C. acknowledges funding by the Simons Foundation (Award No. 568762), the Brinson Foundation, and the National Science Foundation (via Grants No. PHY-2011961 and No. PHY-2011968). Some of our algebraic work used the package {\sc xAct}~\cite{xact} for Mathematica.

\appendix

\section{Chandrasekhar transformation for \texorpdfstring{\(\psi_4\)}{psi4}}
\label{app:psi4_chandra}

In the section, we summarize the derivation of the Chandrasekhar transformation for \(\psi_4\).
We use the same background tetrad as in Eq.~\eqref{eq:bg_tetrad}, the same
Fourier convention \(e^{-i\omega t}\), and the Newman-Penrose Weyl scalar of Refs.~\cite{Newman:1961qr,Chandrasekhar_1983}, written here with the sign appropriate to the signature adopted in Sec.~\ref{sec:introduction},
\begin{equation}
\psi_{4}
:=
C_{\mu \nu \rho \sigma}\,n^{\mu}\bar m^{\nu}n^{\rho}\bar m^{\sigma}.
\label{eq:app_psi4_def}
\end{equation}
This is the convention used throughout, including in
Eq.~\eqref{eq:psi0_def}; \(\psi_{4}\) is obtained from \(\psi_{0}\) by the
tetrad exchange \(l\leftrightarrow n\), \(m\leftrightarrow\bar m\), and
therefore has spin weight \(-2\). 
Along with Eqs.~\eqref{eq:lambda_def}, we further define
\begin{equation}
    D=6M+\lambda r ,
\end{equation}
and note the identity \(\lambda \Lambda=(\ell-1)\ell(\ell+1)(\ell+2)\), used
repeatedly below.


As in Eq.~\eqref{eq:psi0_sep}, the harmonic decomposition of \(\psi_4\)
factorizes into an angular part times a radial coefficient, with the
even- and odd-parity channels carrying independent Fourier frequencies
\(\omega_E\) and \(\omega_O\),
\begin{align}
\psi_4(t,r,\theta,\phi)
&=
A_{4E}^{\ell m}(\theta,\phi)\,e^{-i\omega_E t}\,\psi_4^{E}(r;\omega_E)
\nonumber \\
&+
A_{4O}^{\ell m}(\theta,\phi)\,e^{-i\omega_O t}\,\psi_4^{O}(r;\omega_O),
\label{eq:app_psi4_sep}
\end{align}    
%
Carrying out the tetrad contraction and substituting the axial harmonics given in Eq.~\eqref{eq:vector_sph_harm}, the angular coefficients are
\begin{subequations}\label{eq:app_psi4_angular}
\begin{align}
A_{4E}^{\ell m}
&=
-\bar\eth^{2}Y_{\ell m}
=
-\sqrt{(\ell-1)\ell(\ell+1)(\ell+2)}\;{}_{-2}Y_{\ell m},
\\[4pt]
A_{4O}^{\ell m}
&=
-2i\,\bar\eth^{2}Y_{\ell m}
=
-2i\sqrt{(\ell-1)\ell(\ell+1)(\ell+2)}\;{}_{-2}Y_{\ell m},
\end{align}
\end{subequations}
with \(\bar\eth\) the spin-lowering operator,
\(\bar\eth\,\eta_{s}=-(\partial_\theta-i\csc\theta\,\partial_\phi
+s\cot\theta)\eta_{s}\). These are the spin-weight \(-2\) counterparts of the
factors \(A_{\rm even}^{\ell m}= - \eth^{2}Y_{\ell m}\) and
\(A_{\rm odd}^{\ell m}=2i\,\eth^{2}Y_{\ell m}\) appearing in the
\(\psi_{0}\) map, Eq.~\eqref{eq:chandra_full}, and reduce to them under
\(i\to-i\), as required by the \(m\leftrightarrow\bar m\) exchange. Projecting
Eq.~\eqref{eq:app_psi4_sep} onto \({}_{-2}\bar Y_{\ell_1 m_1}\) and
integrating over the 2-sphere eliminates the angular dependence and leaves a
relation in \((t,r)\) alone. All radial coefficients below are understood
after this projection.

With these conventions the direct RW-gauge projection gives
\begin{align}
\psi_{4}^{E}
&=
\frac{1}{16r^4}
\left[
r^2H_0
+(2M-r)\left(2rH_1+(2M-r)H_2\right)
\right],
\label{eq:app_psi4_even_metric}
\\
\psi_{4}^{O}
&=
\frac{1}{16r^4}
\left[
(2M-i r^2\omega_O)h_0
+(2M-r)
\right. 
\nonumber \\
&\left. 
\times 
\left(
-i r\omega_O h_1
+r h_0'
+(2M-r)h_1'
\right)
\right].
\label{eq:app_psi4_odd_metric}
\end{align}
The rest of the derivation is the elimination of the metric amplitudes in
favor of the ZM and RW master functions.

For the polar sector, the reconstruction maps used in the calculation are given by Eqs.~\eqref{eq:H0dcs}-\eqref{eq:Kdcs}. 
We have verified that the resulting \((H_0,H_1,H_2,K)\) satisfy every linearized even-parity vacuum equation identically once \(\Psi_{\rm ZM}\) obeys the Zerilli equation. The companion relation \(\partial_r H_0=f(r)\,\partial_r K-i\omega_E H_1\), which follows from the \(r\theta\)- component, is a useful consistency check. Following a similar procedure as shown in Sec.~\ref{sec:evenchandra}, we obtain 
\begin{widetext}
\begin{align}
\psi_4^E
&=
C_0^E(r,\omega_E)\Psi_{\rm ZM}(r)
+
C_1^E(r,\omega_E)\Psi_{\rm ZM}'(r),
\label{eq:app_psi4_even_final}
\\
C_0^E
&=
\frac{1}{8r^5D^2}
\Bigg[
-\mathcal A_E\mathcal B_E
+2r\mathcal C_E
\left\{
\Lambda(M-ir^2\omega_E)
-2r\omega_E\left(3iM+r(-i+r\omega_E)\right)
\right\}
\Bigg],
\\
C_1^E
&=
\frac{r-2M}{4r^4D}
\left[
-6M^2
+\lambda r^2(1+ir\omega_E)
+3Mr(2-\Lambda+2ir\omega_E)
\right],
\\
\mathcal A_E
&=24M^2+6\lambda Mr+\lambda \Lambda r^2,
\qquad
\mathcal B_E
=6M^2+2(\Lambda-3)Mr+r^2(2-\Lambda+2r^2\omega_E^2),
\\
\mathcal C_E
&=6M^2+3\lambda Mr-\lambda r^2 .
\end{align}

For the axial sector, the metric amplitudes are reduced using
Eq.~\eqref{eq:dCSRWnorot} together with Eqs.~\eqref{eq:h0_from_h1}-\eqref{eq:h1_from_RW},
which gives the odd-parity Chandrasekhar-reduced expression
\begin{align}
\psi_4^O
&=
C_0^O(r,\omega_O)\Psi_{\rm RW}(r)
+C_1^O(r,\omega_O)\Psi_{\rm RW}'(r)
+C_R^O(r)R_{\ell m}(r),
\label{eq:app_psi4_odd_final}
\\
C_0^O
&=
-\frac{i}{16r^5\omega_O}
\left[
12M^2
+2Mr(-3-\Lambda-3ir\omega_O)
+r^2\left\{\Lambda-2r\omega_O(-i+r\omega_O)\right\}
\right],
\\
C_1^O
&=
-\frac{i(2M-r)}{8r^4\omega_O}
\left[
3M+r(-1-ir\omega_O)
\right],
\qquad
C_R^O
=
-\frac{3M(2M-r)\alpha}{8\kappa r^7}
=
-\frac{6\pi M(2M-r)\alpha}{r^7},
\end{align}
\end{widetext}
where $\kappa=(16\pi)^{-1}$. The scalar term in Eq.~\eqref{eq:app_psi4_odd_final} originates in analogy to its counterpart in the \(\psi_{0}\) map: it is generated when the
second radial derivative \(\Psi_{\rm RW}''\) is eliminated using the
dCS-modified axial equation, Eq.~\eqref{eq:dCSRWnorot}. 

Equations~\eqref{eq:app_psi4_even_final} and~\eqref{eq:app_psi4_odd_final} are
the final nonrotating \(\psi_4\) Chandrasekhar maps. The even
sector is the GR Zerilli reconstruction, while the odd sector contains the
order-reduced dCS scalar source. Setting \(\alpha=0\) recovers the GR
RW result.

\section{Reduced \texorpdfstring{\(\psi_0\)}{psi0}--\texorpdfstring{\(\psi_4\)}{psi4} consistency check}
\label{app:psi0_psi4_reduced_check}

We do not assume, or derive, a Teukolsky-Starobinsky identity in dCS gravity. The usual Teukolsky-Starobinsky identities are fourth-order differential relations between the spin-weight 
\(+2\) and spin-weight \(-2\) Teukolsky variables in GR~\cite{Starobinsky:1973aij,Starobinsky:1974nkd,Press:1973zz,Teukolsky:1974yv}. Since the dCS axial gravitational sector is coupled to the scalar perturbation, a genuine dCS version of the Teukolsky-Starobinsky identity would have to be formulated as a coupled or sourced identity. 
Instead, we present the comparison below only as a consistency check of two independently reconstructed curvature scalars after both have been expressed in the same RW/ZM basis. 

Define
\begin{align}
F(r)=\frac{(r-2M)^2}{4r^2}\,,
\end{align} 
and recall from Eqs.~\eqref{eq:chandra_full} and \eqref{eq:app_psi4_sep} that the radial coefficients \(\delta\Psi_{0,\ell m}^{\rm (even)}(r;\omega_E)\), \(\delta\Psi_{0,\ell m}^{\rm (odd)}(r;\omega_O)\), \(\psi_4^{E}(r;\omega_E)\) and \(\psi_4^{O}(r;\omega_O)\) are already time- and angle-stripped: each is obtained from the full separated field by projecting onto the appropriate spin-weighted harmonic, dividing out the known angular normalization (\(A_{\rm even}^{\ell m}\), \(A_{\rm odd}^{\ell m}\), \(A_{4E}^{\ell m}\) or \(A_{4O}^{\ell m}\)), and removing the factor \(e^{-i\omega_P t}\). For compactness we write, for \(P=E,O\) and \(s=0,4\),
\begin{equation}
\psi_s^{E}\equiv\delta\Psi_{s,\ell m}^{\rm (even)},
\qquad
\psi_s^{O}\equiv\delta\Psi_{s,\ell m}^{\rm (odd)},
\qquad
\psi_s^{O}=\psi_{s,g}^{O}+\psi_{s,R}^{O},
\end{equation}
the last equation splitting the odd radial coefficient into its gravitational RW part and its scalar-source part.

Because these are radial \emph{coefficients} and not radial \emph{solutions}, the frequency appearing below acts only on the explicit frequency-dependent coefficient functions, holding \(\Psi_{\rm RW}(r)\),
\(\Psi_{\rm ZM}(r)\) and \(R_{\ell m}(r)\) fixed. We denote it by
\begin{equation}
\mathcal{R}\!\left[\psi_s^{P}\right](r;\omega_P)
\equiv
\psi_s^{P}(r;-\omega_P) ,
\label{eq:coeff_reflection}
\end{equation}
Direct coefficient comparison of the explicit Chandrasekhar-reduced expressions then gives the parity-split relations
\begin{subequations}
\label{eq:dCS_reduced_psi0_psi4_check}
\begin{align}
\psi_4^{E}(r;\omega_E)
&=
F(r)\,\mathcal{R}\!\left[\psi_0^{E}\right](r;\omega_E),
\label{eq:dCS_reduced_psi0_psi4_even}
\\
\psi_{4,g}^{O}(r;\omega_O)
&=
\,F(r)\,\mathcal{R}\!\left[\psi_{0,g}^{O}\right](r;\omega_O),
\label{eq:dCS_reduced_psi0_psi4_odd_grav}
\\
\psi_{4,R}^{O}(r)
&=
-F(r)\,\psi_{0,R}^{O}(r).
\label{eq:dCS_reduced_psi0_psi4_odd_scalar}
\end{align}
\end{subequations}
The angular normalizations \(A_{\rm even}^{\ell m}\), \(A_{\rm odd}^{\ell m}\), \(A_{4E}^{\ell m}\) and \(A_{4O}^{\ell m}\) have been divided out on both sides of Eq.~\eqref{eq:dCS_reduced_psi0_psi4_check}. 

The scalar part can be checked explicitly. In \(\psi_0\), the scalar contribution is
\begin{equation}
\psi_{0,R}^{O}
=
\frac{3M\alpha}{2\kappa r^5(2M-r)}R_{\ell m} .
\end{equation}
Multiplication by \(F(r)\) gives
\begin{equation}
F(r)\,\psi_{0,R}^{O}(r)
=
\frac{3M\alpha(2M-r)}{8\kappa r^7}R_{\ell m}
=
-\psi_{4,R}^{O}(r).
\label{eq:dCS_scalar_reduced_spin_flip}
\end{equation}
Thus the reduced comparison is a coefficient-level consistency check, not a new local scalar-to-tensor reconstruction rule and not a homogeneous dCS Teukolsky-Starobinsky identity.

\section{Vanishing of the scalar source in the \texorpdfstring{\(\theta\phi\)}{theta-phi} component}
\label{app:ctensor_thetaphi}

Two statements underlie the scalar-free form of Eq.~\eqref{eq:34eqdcsNR}.
First, \(T^{\vartheta}_{\mu\nu}\) is quadratic in \(\nabla_{\mu}\vartheta\), so it
vanishes identically at linear order about a vanishing background scalar.
Second, for the linearized C-tensor: the first term,
\(\nabla_\sigma\vartheta\,\epsilon^{\sigma\delta\alpha(\mu}\nabla_\alpha
R^{\nu)}{}_{\delta}\), vanishes because Schwarzschild is Ricci flat. The
remaining term, \(\nabla_\sigma\nabla_\delta\vartheta\,
{}^{*}R^{\delta(\mu\nu)\sigma}\), contracts \(\nabla_\sigma\nabla_\delta\delta\vartheta\)
against the dual Weyl tensor, which for Schwarzschild BH pairs the \(tr\) block
with the \(\theta\phi\) block. The linear C-tensor therefore has support only
on the mixed \(tr\times\theta\phi\) components. Explicitly, with Eq.~\eqref{eq:sfperts},
the only non-vanishing components are
\begin{subequations}
\begin{align}
\delta C_{tA}&=\frac{3M(r-2M)}{r^{6}}
\left(2R_{\ell m}-r\partial_r R_{\ell m}\right)S_A\,e^{-i\omega t},\\
\delta C_{rA}&=\frac{3iM\omega}{r^{3}(r-2M)}\,R_{\ell m}\,S_A\,e^{-i\omega t},
\qquad A\in\{\theta,\phi\},
\end{align}
\end{subequations}
All purely angular components vanish,
\begin{equation}
\delta C_{AB}[\delta\vartheta]=0,
\qquad A,B\in\{\theta,\phi\},
\label{eq:Cthetaphi_vanishes}
\end{equation}
and in particular \(\delta C_{\theta\phi}[\delta\vartheta]=0\), for every \(\ell\).
The scalar coupling is therefore confined to the mixed \(tA\) and \(rA\)
equations. It enters both the axial master equation and the finite-radius
sourced RW-to-CPM constraint of
Eq.~\eqref{eq:rw_cpm_time_dictionary_sourced}, but never the \(\theta\phi\)
reconstruction equation  given in Eq.~\eqref{eq:34eqdcsNR} used for the leading tensor strain.

\bibliographystyle{apsrev4-1}
\bibliography{reference_new}

\end{document}